\documentclass[useAMS,usegraphicx,usenatbib,a4paper]{mn2e}
\usepackage[]{amsmath,amssymb}
\usepackage{mathrsfs}
\usepackage{aas_macros}
\usepackage{color}
\usepackage[dvips, bookmarks, pdfborder={0 0 0.1}]{hyperref}
\usepackage{mn2e-bst} 
\usepackage{cuted}

\newcommand{\D}{\mathrm{d}}

\newcommand{\msunh}{h^{-1}\mathrm{M}_\odot}
\newcommand{\msun}{\mathrm{M}_\odot}
\graphicspath{{figures/}}

\newcommand{\myplot}[1]{\includegraphics[width=0.5\textwidth]{#1}}
\newcommand{\myplottwo}[2]{\myplot{#1}\myplot{#2}}

\newcommand{\mytab}{\begin{table}[htb]}
\newcommand{\myfig}{\begin{figure}[htbp]}

\newcommand{\mybibstyle}{mymn}

\DeclareMathOperator\erf{erf}
\usepackage{xspace}
\newcommand{\hl}[1]{\color{black}#1\color{black}\xspace}
\begin{document}
\title[subhalo distribution]{A unified model for the spatial and mass distribution of subhaloes}
\author[J. Han et al.]{Jiaxin Han,$^{1}$\thanks{jiaxin.han@durham.ac.uk} 
Shaun Cole,$^1$ Carlos S. Frenk,$^1$ Yipeng Jing$^{2,3}$\\
$^1$Institute for Computational Cosmology, Department of Physics,
Durham University, South Road, Durham, DH1 3LE\\
$^2$Centre for Astronomy and Astrophysics, Department of Physics and Astronomy, Shanghai Jiao Tong University, Shanghai 200240, China\\
$^3$IFSA Collaborative Innovation Centre, Shanghai Jiao Tong University, Shanghai, 200240, China\\
}
\maketitle

\begin{abstract} 
N-body simulations suggest that the substructures that survive inside dark matter haloes follow universal distributions in mass and radial number density. We demonstrate that a simple analytical model can explain these subhalo distributions as resulting from tidal stripping which increasingly reduces the mass of subhaloes with decreasing halo-centric distance. As a starting point, the spatial distribution of subhaloes of any given infall mass is shown to be largely indistinguishable from the overall mass distribution of the host halo. Using a physically motivated statistical description of the amount of mass stripped from individual subhaloes, the model fully describes the joint distribution of subhaloes in final mass, infall mass and radius. As a result, it can be used to predict several derived distributions involving combinations of these quantities including, but not limited to, the universal subhalo mass function, the subhalo spatial distribution, the gravitational lensing profile, the dark matter annihilation radiation profile and boost factor. This model clarifies a common confusion when comparing the spatial distributions of galaxies and subhaloes, the so called ``anti-bias'', as a simple selection effect. We provide a \textsc{Python} code \textsc{SubGen} for populating haloes with subhaloes at \url{http://icc.dur.ac.uk/data/}.
\end{abstract} 


\section{Introduction}
With advances in numerical simulations, the statistics of
subhaloes inside dark matter haloes are being quantified to ever higher 
accuracy. These statistics cover a variety of different
properties of subhaloes, including mass, position, size,
maximum circular velocity, orbit, mass stripping rate and accretion
time. Among these statistics, two outstandingly simple distributions
are the subhalo mass function and the spatial
distribution. It is well established that the mass function of
subhaloes follows a universal power-law form (except for an exponential
high mass tail), $\D N/\D \ln m=A M_{\rm host} m^{-\alpha}$, with a
universal amplitude, $A$, and a universal slope, $\alpha\simeq 0.9$,
that are both independent of the host halo mass, $M_{\rm
host}$ \citep[e.g.][]{Gao04a,Phoenix,Giocoli08b}. The spatial distribution is
known to be less concentrated than both the DM and galaxy
distributions \citep[see e.g.,][and references therein]{Gao04b,
Libeskind05, Diemand04}. With recent zoom simulations that are able to
resolve subhaloes spanning $\sim7$ orders of magnitude in mass and
$\sim2$ orders in separation from the centre of the host halo, the
spatial distribution of subhaloes is also found to be universal, that is
independent of the mass of subhalo \citep{Aquarius,Phoenix,CoCo}. 

Tracking progress in simulations are models for the formation and
evolution of subhaloes. The majority of the models are
semi-analytical~\citep[e.g.,][]{TB01,Benson02,TB04,TB05a,TB05b,Zentner05,PB05},
typically starting from Monte-Carlo merger trees to evolve subhaloes
dynamically after infall. Each subhalo is assigned a set of initial
orbital parameters according to distributions extracted from
simulations. The orbits of these subhaloes are then evolved inside the
host halo, with the mass of the subhalo updated at each timestep
according to the tidal radius and a dynamical timescale. The
statistical properties of the final subhalo population, such as the mass
and velocity functions and the spatial distribution can be obtained. These
models typically involve a number of free parameters that are
calibrated by comparing the predicted subhalo distributions to
simulations. The advantages of such models are that they can
incorporate detailed physical processes, including dynamical
friction, tidal heating, tidal stripping and total disruption. Furthermore, these
models are often incorporated in semi-analytic models of
galaxy formation. However, being a semi-analytical model means
one has to resort to Monte-Carlo realizations of merger trees and
detailed evolution of individual orbits to obtain population
statistics. 

A simplification was introduced by \citet{vdB05}, in which
they only considered the average mass loss rate of subhaloes without
evolving individual orbits. With information on the infall mass
and infall time of subhaloes obtained from Monte-Carlo merger trees,
they can evolve the overall mass distribution of subhaloes to recover
the final mass and velocity functions~\citep{Jiang14,vdB14}. However,
the price of not following individual orbits is the loss of spatial
information, so that these models cannot predict the spatial
distribution of subhaloes.

In this work, we propose a simple and fully analytical model that
simultaneously predicts both the mass function and spatial
distribution of subhaloes, focusing on the key ingredients that shape
these distributions. We start from empirical results on the infall
mass function and spatial distribution of subhaloes labelled by their
infall masses (i.e., mass at accretion), which are found to be
simple to describe and easy to interpret. The only difference
between these distributions and the final distributions is reduction
in mass of each subhalo.  For this reason, we will call the
distribution labelled by infall mass as ``unevolved'' and
that labelled by final mass as ``evolved''. To obtain the
latter, we only need to specify the connection between the final mass
and the infall mass of each subhalo. This is achieved by a physically
motivated statistical description of final-to-infall mass ratio at
each halo-centric radius, rather than by relying on recipes to evolve the
mass and orbit of each subhalo individually.

The idea that the mass and spatial distribution can be derived from
coupling the unevolved distributions with subhalo
stripping was already explored by \citet{Lee04} and
\citet{OguriLee04}. An extended Press-Schechter~\citep[hereafter EPS;][]{BBKS,Bond91,LC93,SMT01} approach was adopted to predict
the progenitor mass function at each radius inside the host. The mass
and location of these progenitors were then adjusted following simple
tidal stripping and dynamical friction predictions. The resulting mass
and spatial distributions of subhaloes largely agreed with the results
from N-body simulation available at that time. These models are
fully analytical, with only one free parameter in \citet{Lee04} and
essentially zero free parameters in the improved
version~\citep{OguriLee04}. However, the model assumptions are not
directly validated by simulations, and some of them are too idealized
to be realistic. For example, subhaloes are all placed on circular
orbits, which is certainly not the case in cosmological simulations. They also
assume that initially the subhalo spatial distribution traces the density
profile of the host halo. This differs 
from the unevolved spatial distribution 
(the distribution of subhaloes selected by their infall mass)
in that the former describes the
location of the progenitor haloes at the formation time, while the latter
describes the location at the final time. However, the circular orbit
assumption, coupled with the assumed initial spatial distribution, leads
to the same unevolved spatial distribution as we find in this
work. This could explain why they are largely able to reproduce the subhalo
spatial distribution in simulations despite unrealistic
assumptions. In contrast to their models which mostly start from
theoretical assumptions, our model is built upon empirically validated
assumptions. The success of our model thus serves as a promising
starting point for more realistic first-principle models that attempt to
understand fully the formation and evolution of subhaloes.

We use two sets of high-resolution zoom-in simulations (the
Aquarius project, \citealp{Aquarius}, and the Phoenix project, \citep{Phoenix})
to verify and calibrate our model. The high resolution of these
simulations enables us to make a detailed statistical analysis of the
spatial distribution and stripping of subhaloes with unprecedented
precision. Each halo is also simulated at a series of resolutions,
allowing us to distinguish physical properties from numerical
artefacts by carrying out convergence studies.

With the calibrated assumptions, the model simultaneously
recovers both the final subhalo mass function and spatial distribution accurately.
 The full model specifies analytically the joint distribution of
subhalo infall mass, final mass and location inside the host.
 Such a joint distribution is ideal to obtain fast
realizations and for Halo Occupation Distribution
(HOD) modelling of subhaloes inside a host halo. We
provide a \textsc{Python} implementation, \textsc{SubGen}, for such
purposes. We also present several example applications, including the 
implication to the universality of the subhalo mass function, a
prediction for lensing measurements of subhalo masses and the
modelling of dark matter annihilation radiation including the 
boost factor from subhaloes. 

This paper is organized as follows: the simulations used to verify and
calibrate the model are introduced in Section~\ref{sec:simu}; the key
idea and the effectiveness of our model is briefly demonstrated
using a simplified version in Section~\ref{sec:simplemodel}; the
three model assumptions are then investigated and validated in detail
in Section~\ref{sec:components}; the full statistical model is
presented and extended to arbitrary
host masses in Section~\ref{sec:fullmodel}; some implications and applications of the model are
discussed in Section~\ref{sec:applications}; finally we summarize and
conclude in Section~\ref{sec:summary}.

In most cases, we use lower case to refer to properties of the
subhaloes, and capitals to refer to the properties of the host. For
example, $m$ and $r$ refer to the mass and radius of a subhalo, and
$M$ and $R$ refer to the mass of the host and the radial location
within a host. The virial mass, $M_{200}$, is defined to be the mass
inside the virial radius, $R_{200}$, inside which the average density
equals 200 times the critical density of the universe. We use a
mass scale $m_0=10^{10}\msunh$ to normalize the halo mass in
scaling relations.

\section{Simulations}\label{sec:simu}
We use two sets of high resolution simulations to verify and
calibrate our model assumptions. One is Aquarius \citep{Aquarius}, a set of zoomed-in
simulations of Milky Way sized haloes each with a mass of $\sim
10^{12}\msunh$. Six different haloes are simulated (named A to F) in
total, at five different levels of resolutions (labelled 1 to 5 from
high to low resolution, with particle masses ranging from
$\sim10^3\msunh$ to $\sim10^6\msunh$). The other is the Phoenix \citep{Phoenix}
simulations of nine cluster sized haloes each with a mass of $\sim
10^{15}\msunh$, run at four different levels of resolutions (labelled
1 to 4, with particle masses ranging from $6\times10^5\msunh$ to $10^{8}\msunh$, so that the same level number corresponds to a
similar number of particles inside the host halo both in the Aquarius and
Phoenix simulations). For demonstartion purposes we will mainly 
focus on Aquarius halo A (named AqA1 to AqA5 from
level 1 to 5). 
The properties studied in this work behave
qualitatively similar across the different haloes, with only small
differences in parameter values. 
 Both simulations are run with cosmological parameters
$\Omega_{\rm m}=0.25$, $\Omega_\Lambda=0.75$, $\sigma_8=0.9$, $n_{\rm s}=1$
and $h=0.73$, where $H_0=100 h \,{\rm km\,s^{-1} Mpc^{-1}}$ is the Hubble
constant.

The subhaloes in these simulations are identified using
\textsc{subfind} \citep{subfind}, with merger trees built with the
\textsc{DTrees} algorithm \citep{DHalo}. Most results in this paper are expected to be not
sensitive to the choice of halo finder or tree builder. The mass, $m$,
of a subhalo is defined to be its self-bound mass. The exact
definition of infall could have some ambiguity, depending on how the
boundary of the host halo is defined, and also because a subhalo may
cross the boundary several times during its orbit. To avoid this
ambiguity, we define the infall mass, $m_{\rm acc}$, to be the maximum
bound mass a subhalo ever had in all previous snapshots. Adopting a
different definition could lead to slightly different parameter values
but does not qualitatively affect any of the
assumptions or conclusions of our model. The current position of a
subhalo is defined to be the current position of the most-bound
particle it had when it was last resolved. In this way we follow
both resolved subhaloes and those stripped below the subhalo mass
resolution (defined to be 20 bound particles).

\section{The model in a nutshell}\label{sec:simplemodel}

The basic features of our model can be demonstrated in a simplified
version as follows. Let us assume:
\begin{enumerate}
 \item The unevolved subhalo mass function, i.e., the distribution of
 infall masses, $m_{\rm acc}$, of the subhaloes accreted at all
 previous redshifts, is a power-law function,
 \[
\frac{\D N}{\D \ln m_{\rm acc}}\propto m_{\rm acc}^\alpha,
\] with $\alpha\sim 0.9$ according to previous results~\citep[e.g.][]{Giocoli08b}.
 \item The unevolved spatial distribution of subhaloes, i.e., the
 spatial distribution with a given infall mass,\footnote{This is not
 to be confused with the spatial distribution of subhaloes at the infall
 time. By definition, at the infall time the subhaloes are all located
 near the host halo boundary, while the unevolved spatial
 distribution refers to the distribution of the final positions of
 subhaloes in each infall mass bin.} traces the mass density profile,
 $\rho(R)$, of the host halo,
 \[ 
\frac{\D N(R|m_{\rm acc})}{\D^3 R}\propto \rho(R),
\]
where $\D^3R=4\pi R^2\D R$ is the volume element.
 \item The mass of the subhalo evolves due to tidal stripping after infall. The stripping is stronger at a smaller radius, which can be parametrized as 
 \[
  \frac{m}{m_{\rm acc}}\propto R^\beta.
  \] 
This means the final mass, $m$, is fully determined by $m_{\rm acc}$
and $R$. For isothermal density profiles, tidal stripping predicts
$\beta=1$. For more realistic density profiles such as NFW, we expect
$\beta\sim1$.
\end{enumerate}

\begin{figure*}
 \myplottwo{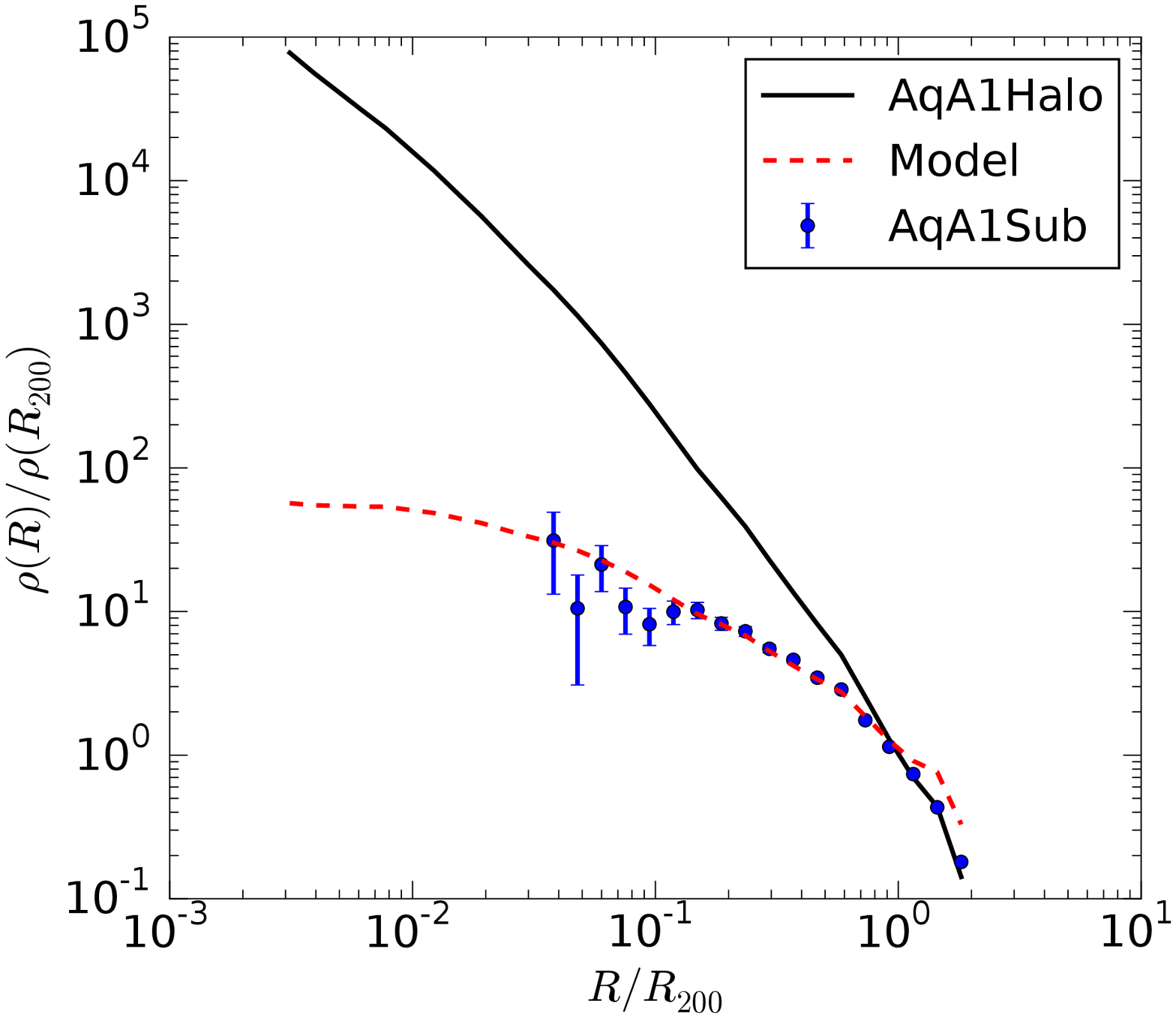}{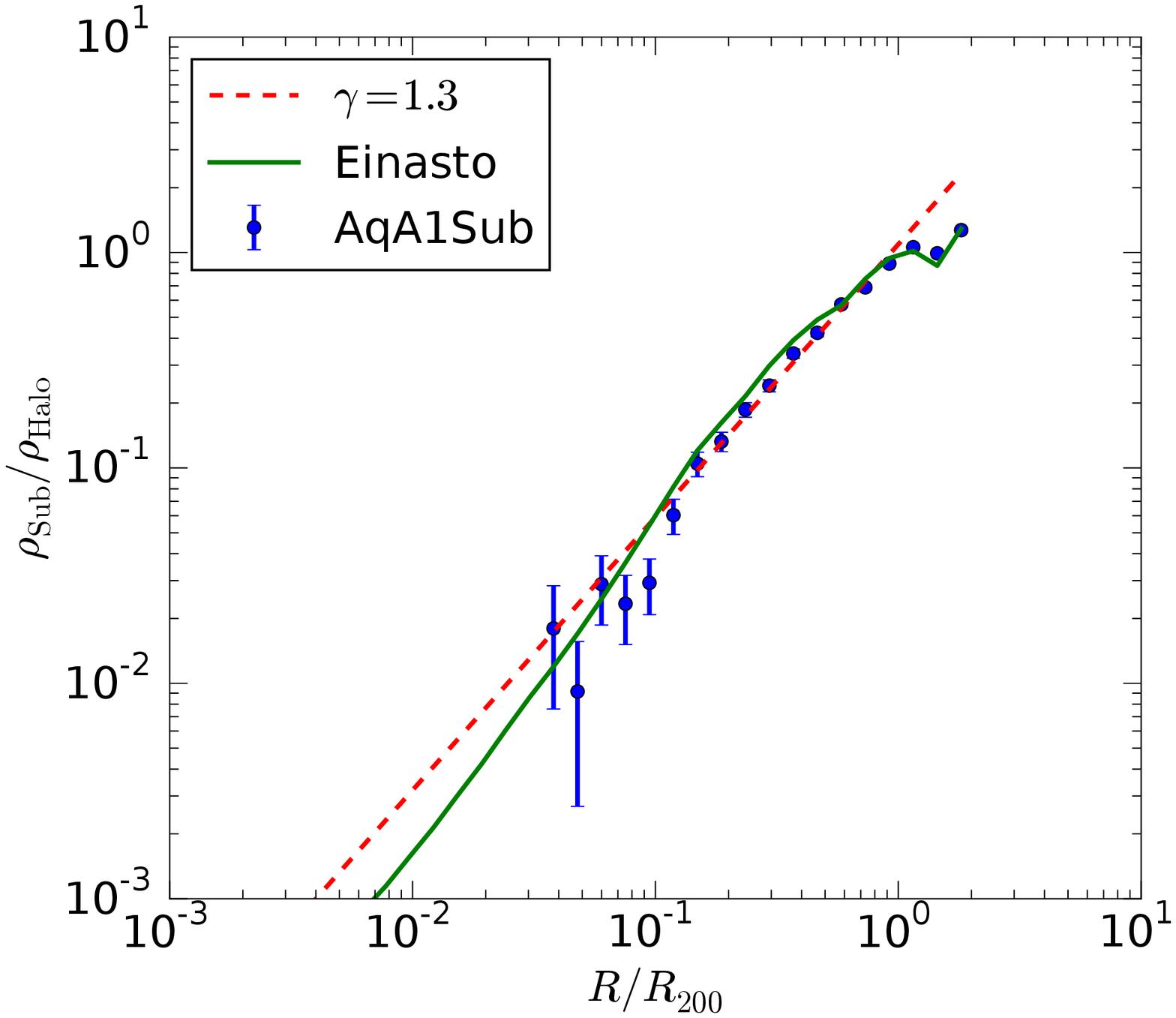}
 \caption{Spatial distribution of subhaloes in AqA1. The data points
 with Poisson errorbars show the number density profile of subhaloes
 resolved with more than 1000 particles (or $10^{-6}M_{200}$, where
 $M_{200}$ is the virial mass of the host halo) at $z=0$. The black
 solid line shows the matter density profile of the host halo. Both
 profiles are normalized by their values at $R_{200}$, the host virial
 radius. Right: ratio between subhalo number density and the DM
 density of the host halo, normalized to unity at $R_{200}$. The
 points are from the same data as in the left. The green solid line
 adopts the best-fit Einasto profile of \citet{Aquarius} for the
 subhalo number density. In both panels, the red dashed line show a
 power-law fit of the form $(R/R_{200})^\gamma$ to $\rho_{\rm
 Sub}/\rho_{\rm Halo}$ inside $R_{200}$. }
\label{fig:prof}
\end{figure*}

Following these assumptions, the evolved distribution is given by
\begin{align}
 \frac{\D N(m,R)}{\D \ln m\D^3R} & = \frac{\D N(m_{\rm acc},R)}{\D \ln m_{\rm acc}\D^3R}\frac{\D \ln m_{\rm acc}}{\D \ln m} \nonumber\\
 &=\frac{\D N(m_{\rm acc})}{\D \ln m_{\rm acc}} \frac{\D \ln m_{\rm acc}}{\D \ln m} \tilde{\rho}(R)\nonumber\\ 
 &\propto m^{-\alpha} R^{\gamma}
 \tilde{\rho}(R),\label{eq:simple_model}
\end{align}  with $\gamma=\alpha\beta\sim 1$. 

Equation~\ref{eq:simple_model} immediately reveals three very
interesting features of the subhalo distribution: 
\begin{enumerate}
 \item The final subhalo mass function shares the same slope as the
 infall mass function.
 \item The final subhalo number density profile is shallower than the
 host halo density profile (or equivalently, the unevolved subhalo
 number density profile), by a factor $R^\gamma$.
 \item The subhalo mass function and spatial distribution are
 separable, meaning that the spatial distribution of different mass
 subhaloes are the same except for a change in amplitude.
\end{enumerate}
These features agree qualitatively with existing results on the
subhalo distribution. In Fig.~\ref{fig:prof}, the predicted spatial
profile is compared with the simulations, focusing on its shape. The
ratio between the subhalo spatial profile and the host halo density
profile indeed agrees very well with a power-law, with $\gamma=1.3$
for Aquarius halo A.

\begin{figure}
 \myplot{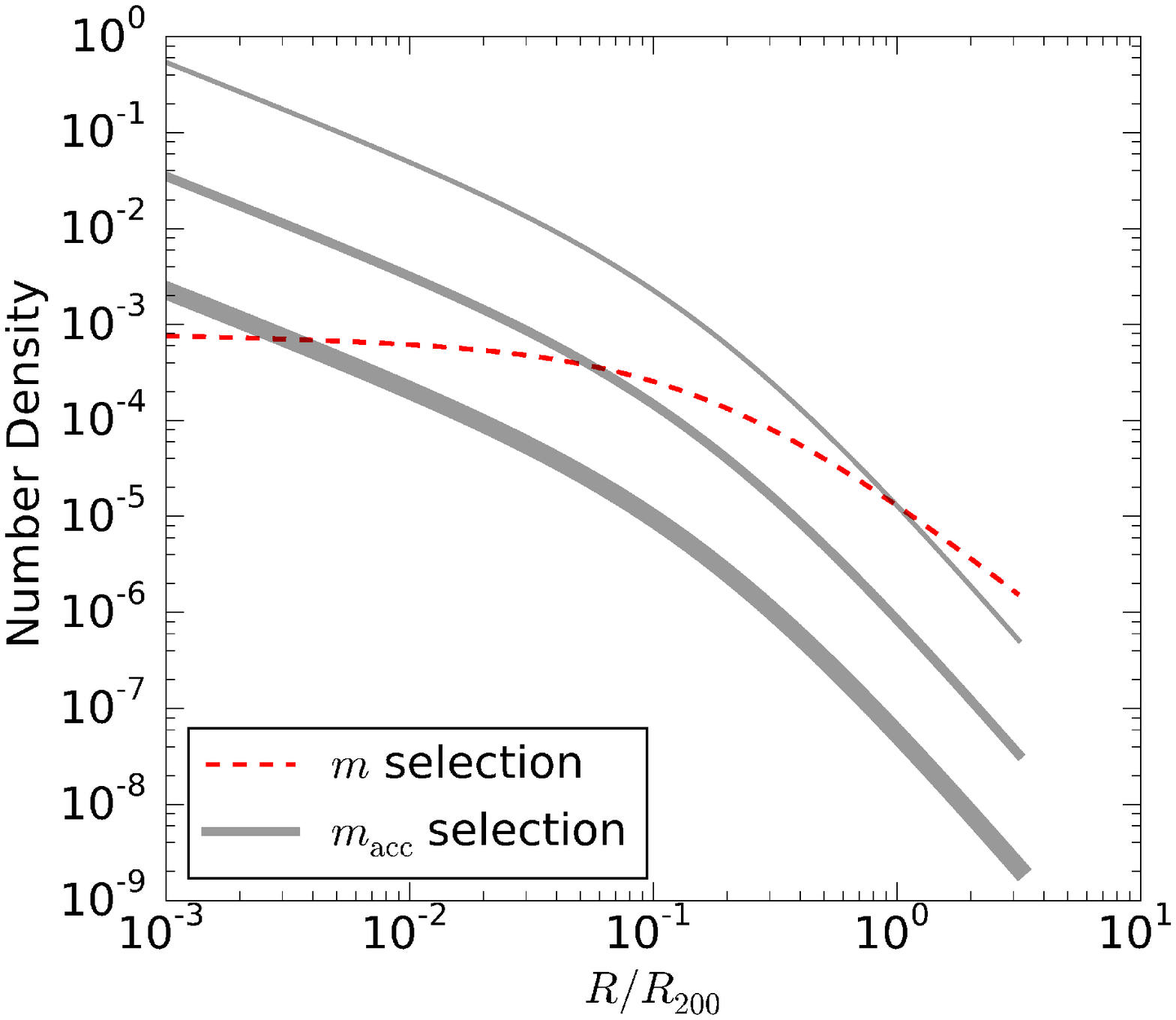}
 \caption{Illustration of the spatial profile of subhaloes. Grey solid
 lines are the spatial profiles of subhaloes with a given infall mass,
 $m_{\rm acc}$, with thicker lines corresponding to larger $m_{\rm
 acc}$. The red dashed line is the profile for subhaloes with a given
 final mass, $m$. For subhaloes with same $m$, those found in the
 inner halo have larger $m_{\rm acc}$ because they are more
 stripped. The number densities of subhaloes selected with different
 $m_{\rm acc}$ at each radius form the profile of the subhaloes with
 the same $m$.}\label{fig:profdemo}
\end{figure}

The difference between the unevolved and evolved subhalo spatial distribution can be understood as the result of a selection function, which is illustrated in Fig.~\ref{fig:profdemo}. According to our first two assumptions, the spatial profiles of subhaloes at fixed infall mass all have the same shape, with more massive subhaloes having a profile with a lower normalization. In the presence of mass stripping, subhaloes selected with the same final mass correspond to populations with different infall mass at different radii. Those found in the inner halo are on average more stripped, so they started with a larger infall mass. Hence the factor $R^\gamma$ by which the profile is suppressed describes how the amplitude of the unevolved spatial profile varies with infall mass for subhaloes selected to have the same final mass at different radii. 

\section{Verifying Model Components}\label{sec:components}
Below we carefully test and adapt the assumptions of the basic model presented above using simulations. We focus on halo A from the Aquarius simulations as our prime example.

\subsection{The spatial PDF of accreted objects}
The density profile of the host halo can be interpreted as the probability distribution function (PDF) for the current position of particles accreted by the halo. As a first approximation, we assume the same PDF applies to each accreted subhalo, regardless of its mass at accretion. This PDF is then given by the normalized density profile of the host halo
\begin{equation}
 \tilde{\rho}(R)=\frac{\rho(R)}{M_{200}},\label{eq:rho_def}
\end{equation} where $\rho(R)$ is the density profile of the host and $M_{200}$ is the virial mass of the host halo. Adopting an NFW profile for the host
\begin{equation}
 \tilde{\rho}(R)=\frac{\tilde{\rho}_{\rm s}}{(R/R_{\rm s})(1+R/R_{\rm s})^2},\label{eq:rho}
\end{equation} where $\tilde{\rho}_{\rm s}=1/\int \frac{\D^3 R}{(R/R_{\rm s})(1+R/R_{\rm s})^2}$ is the normalization, and $R_s$ is the scale radius. 

\begin{figure*}
 \myplottwo{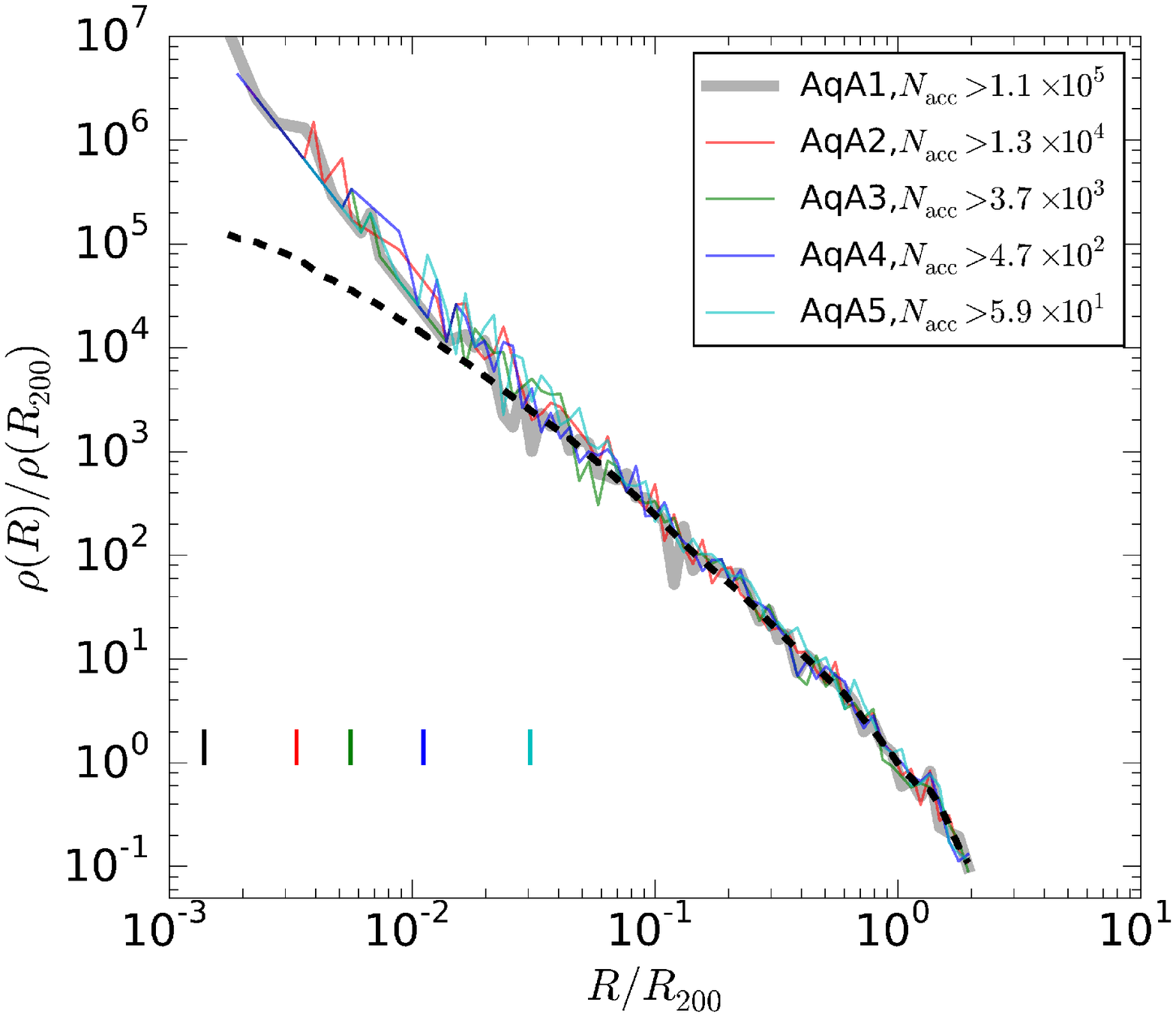}{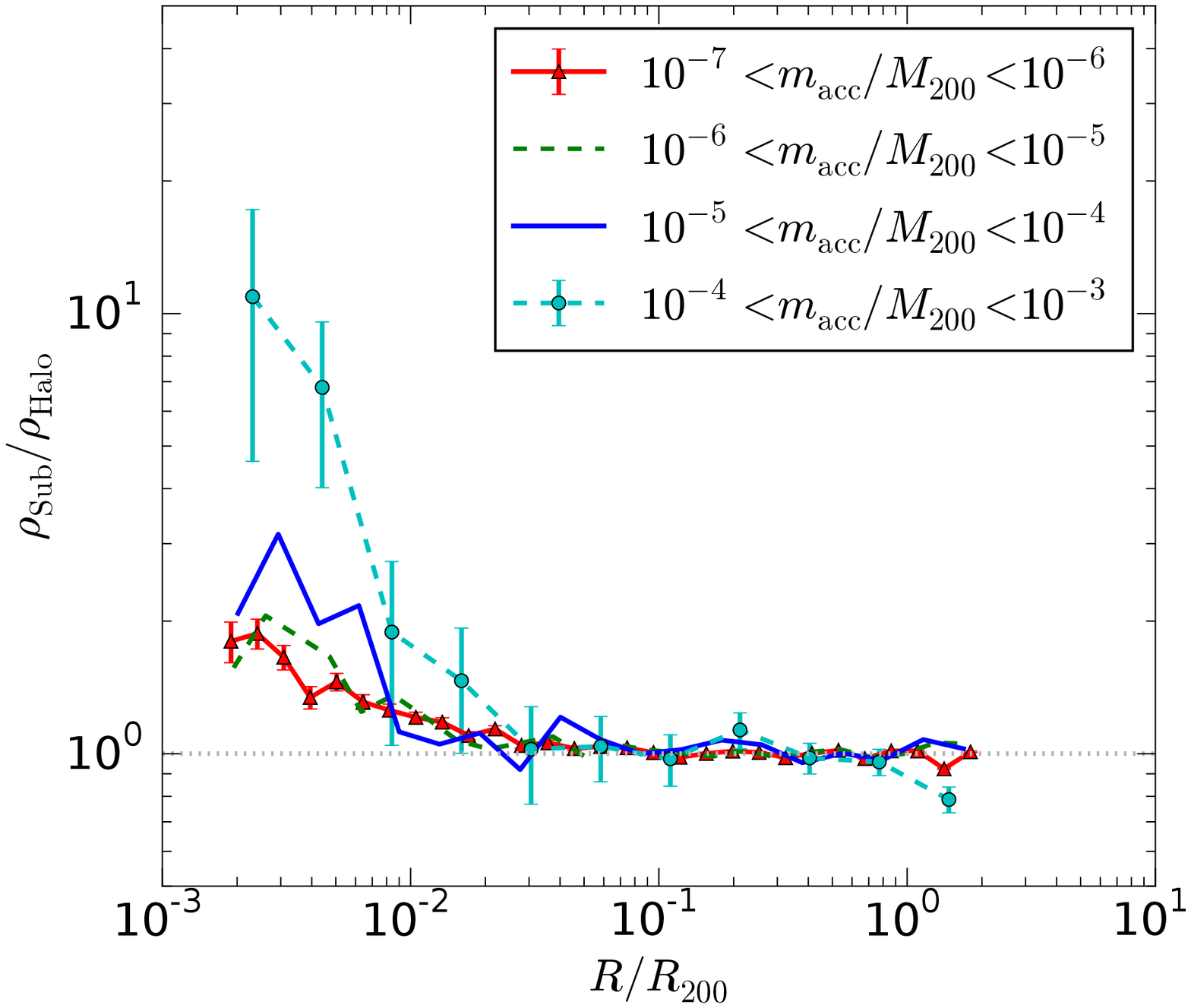}
 \caption{Left: Density profiles of subhaloes in Aquarius halo A with infall mass $m_{\rm acc}>10^{-4}M_{200}$, where $M_{200}$ is the host halo virial mass at $z=0$. The profiles are normalized by their value at $R_{200}$, the host virial radius. Different solid lines correspond to results from different resolution simulations of the same halo (A1 to A5), each selected with the same infall mass threshold. This threshold translates into different threshold in the number of particles at accretion, $N_{\rm acc}$, for each simulation as labelled. For comparison, the dashed line show the density profile of the host halo, similarly normalized by its value at $R_{200}$. The vertical short lines in the bottom mark the convergence radii at the five resolution levels~\citep{AqProf}. Right: Ratio of the subhalo number density profile to the host halo mass density profile in Aquarius halo A1, normalized to unity on large scale. Subhaloes are binned according to their infall mass $m_{\rm acc}$. \hl{Errorbars are shown for the lowest and highest mass bins according to the Poisson noise in the subhalo counts.} The horizontal dotted line marks $\rho_{\rm Sub}/\rho_{\rm Halo}=1$ for reference.}\label{fig:InfallProf}
\end{figure*}

The above assumption is natural if subhaloes follow similar orbits to dark matter particles. In a steady-state halo, the density profile is fully determined by the distribution of orbits of the particles, because the distribution of particles around their orbits is fixed by the travel time at each position \citep{Han15}. If subhaloes follow a similar distribution of orbits to that of the particles, they would naturally form the same spatial profile as that of the dark matter particles. The dynamics of subhaloes differs from that of particles due to dynamical friction. However, for small subhaloes for which dynamical friction is not important, one would expect a quite similar PDF to that of DM particles. Indeed it has been found that subhaloes have a distribution of orbits similar to that of the particles, with a very weak mass dependence \citep[e.g.][]{Lilian}. This means subhaloes are, in terms of their dynamics and spatial distribution, approximately indistinguishable from each other, as well as from dark matter particles, as long as each subhalo is persistently traced. Such a picture can be summarized as ``unbiased accretion'' of subhaloes relative to the accretion of dark matter particles. 

In Fig.~\ref{fig:InfallProf} we test this assumption directly with the Aquarius simulations. We plot the density profile of the host halo and the number density profile of subhaloes, both normalized by their density at the host virial radius $R_{200}$. The subhaloes include all the objects that have been accreted by the host halo at any previous redshift, selected according to their mass at the time of infall, $m_{\rm acc}$. The positions of these objects are defined by the current position of the most-bound particle from when the subhalo was last resolved. Except for a cuspier inner profile due to dynamical friction, the subhaloes closely follow the DM distribution. This is seen more clearly in the right panel, where subhaloes in halo A1 are split into infall mass bins. The subhaloes closely follow the DM profile of the host halo, except for an excess in the very inner part which is more prominent for more massive subhaloes. 
The cuspier inner profile is consistent with expectations from dynamical friction, which causes subhaloes to sink toward the halo centre and is stronger for more massive objects. For the majority of subhaloes which are located in the outer halo and the low mass end, however, this effect is not important.  

\begin{figure}
 \myplot{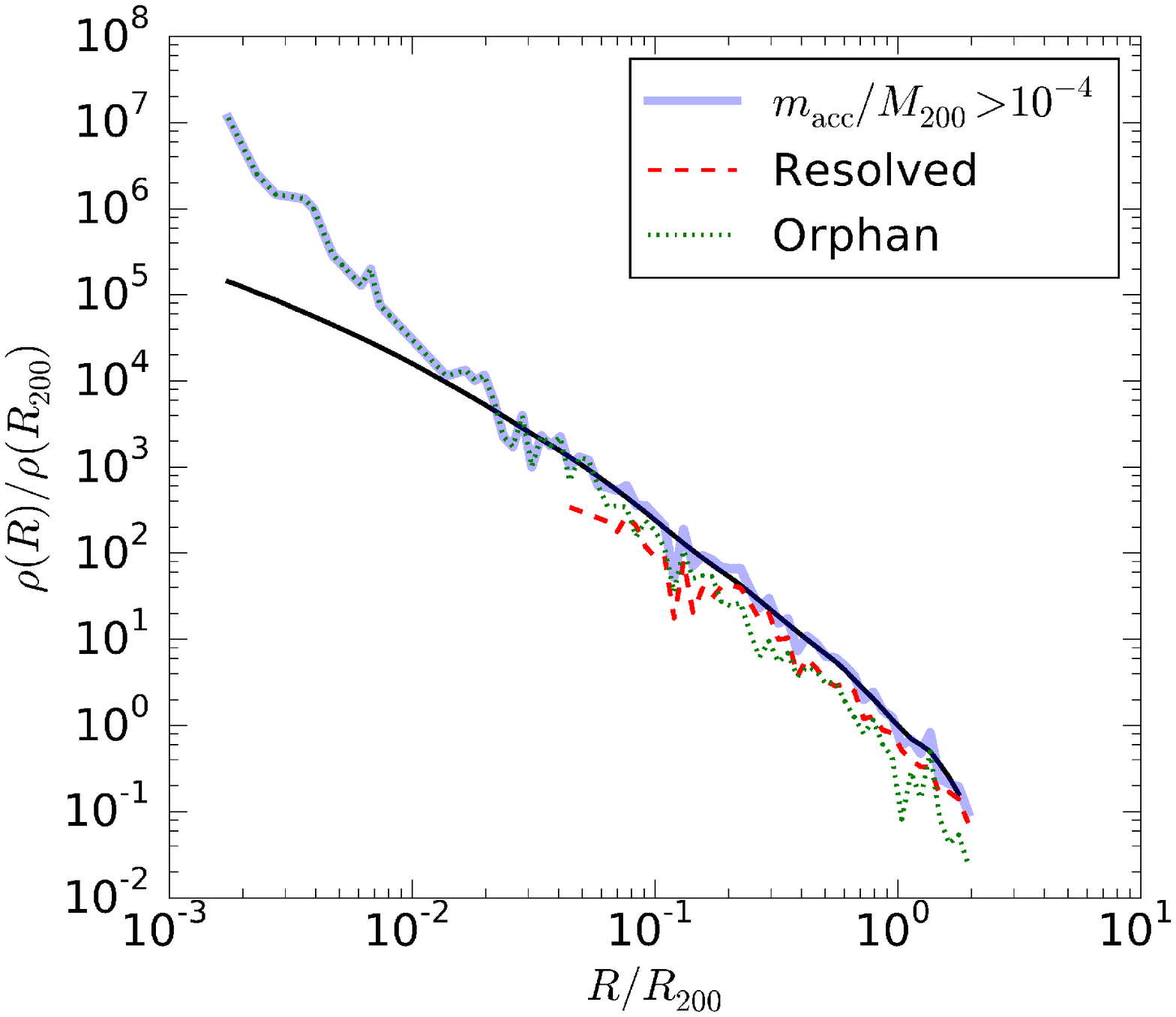}
 \caption{The profile of subhaloes with infall mass $m_{\rm acc}>10^{-4}M_{200}$ in A1, decomposed into resolved and unresolved (orphan) populations at $z=0$. The profiles are normalized by the total subhalo number density at $R_{200}$. For reference, the dark matter density profile normalized to unity at $R_{200}$ is also shown as the thin solid line. }\label{fig:decompose}
\end{figure}

One may question the reliability of using the most bound particle to extract the radial distribution for un-resolved subhaloes. The choice of using the most bound particle to represent unresolved subhaloes is in line with some semi-analytical models~\citep[e.g.][]{subfind,Gao04b}. In these models, orphan galaxies are associated with the most bound particles, and the resulting radial distribution reproduces that of observed galaxies well. It also closely follows that of the DM. 

\begin{figure*}
 \myplottwo{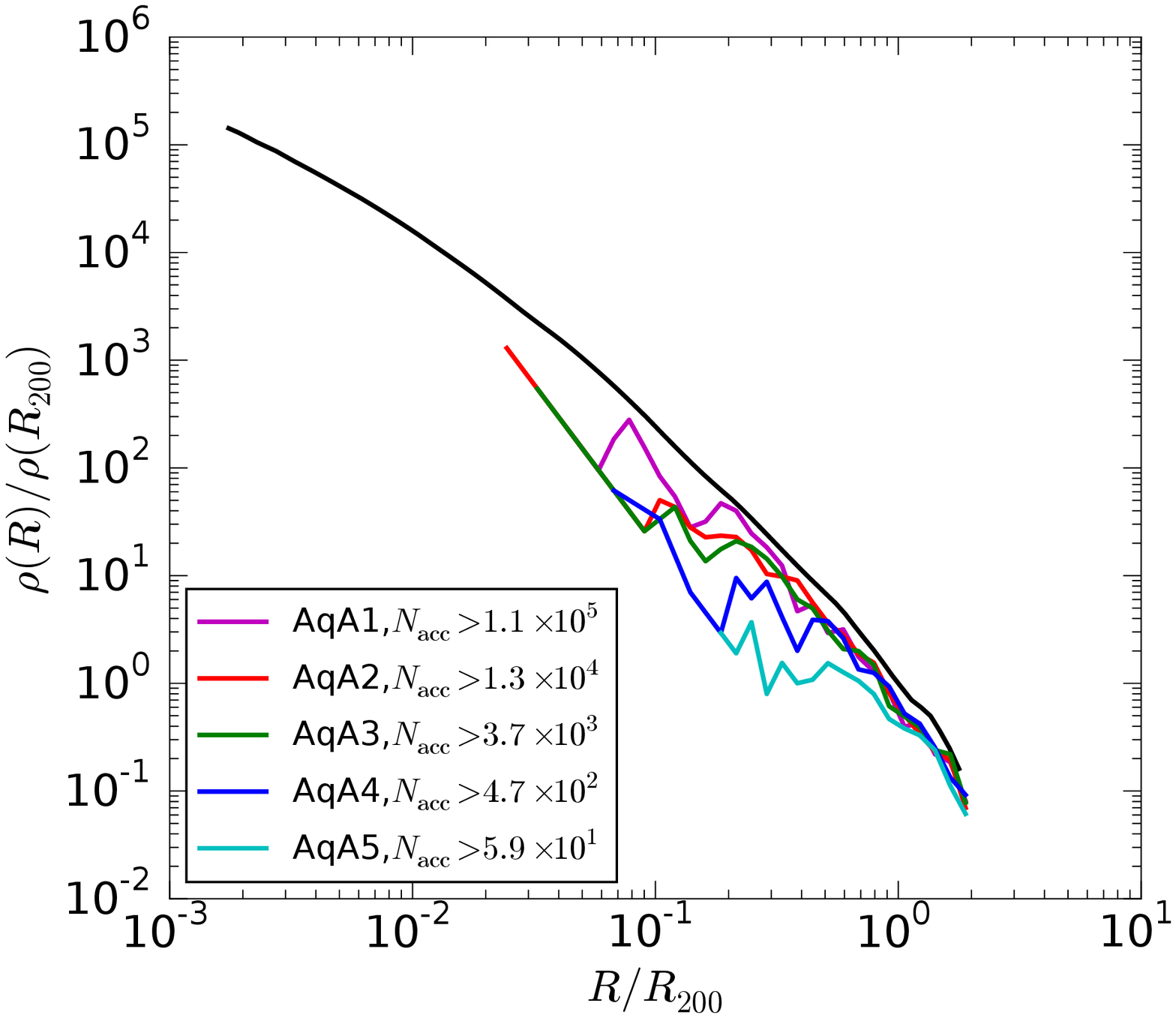}{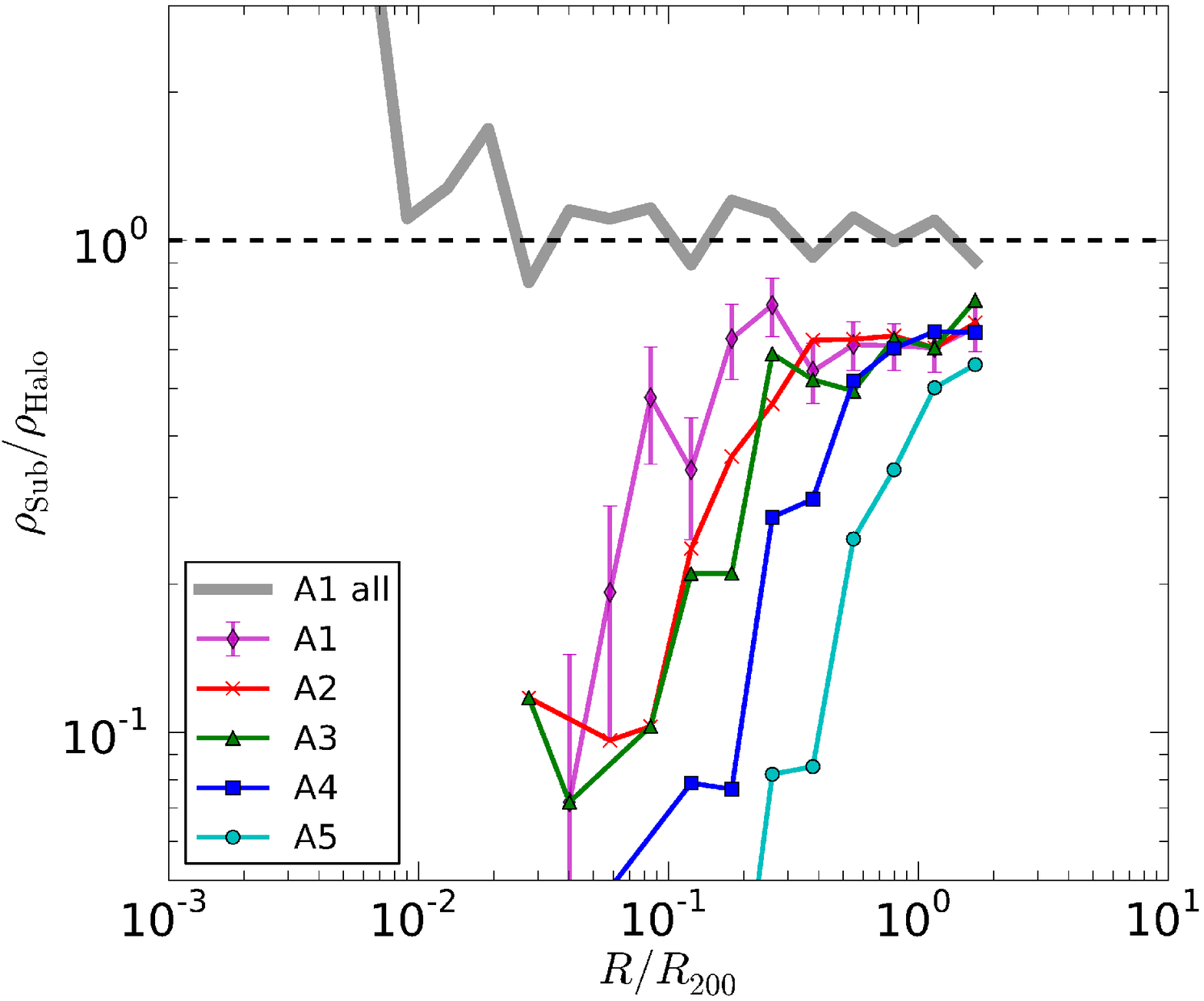}
 \caption{Number density profile for resolved subhaloes. This is similar to Fig.~\ref{fig:prof}, but only for subhaloes that are still resolved at $z=0$, with infall mass $m_{\rm acc}>10^{-4}M_{200}$. Left: subhalo profiles resolved at different resolutions in halo A, commonly normalized by the total subhalo number density at $R_{200}$ in A1. The black solid line is the DM density profile, normalized to unity at $R_{200}$. Right: the ratio between the subhalo profiles and the DM profile as in the left panel. \hl{Poisson errorbars are only shown for the A1 curve for clarity. } The thick solid line is that for all accreted subhaloes in A1. }\label{fig:ResolvedProf}
\end{figure*}

The reliability of tracking unresolved subhaloes with most bound particles can also be checked directly in the simulations. In the left panel of Fig.~\ref{fig:InfallProf} we carry out a convergence study of the unevolved profile. As the resolution of the simulation increases from AqA5 to AqA1 by $\sim 5$ orders of magnitude in particle numbers, more and more unresolved objects become resolved. However, we do not observe any significant change in the unevolved subhalo distribution. Note that even at our highest resolution, only half of the subhaloes in the left panel of Fig.~\ref{fig:InfallProf} remain resolved at $z=0$. These resolved objects dominate at large radii, with a profile that still follows that of DM as seen in Fig.~\ref{fig:decompose}. A convergence study of the spatial distribution of these resolved objects is carried out in Fig.~\ref{fig:ResolvedProf}. As the resolution of the simulation increases, more and more subhaloes can be resolved at $z=0$, and the shape of the subhalo profile approaches that of the DM profile down to smaller and smaller radii. The amplitude of the subhalo profile also converges to $\sim 60\%$ of that of the entire accreted population as seen in the right panel of Fig.~\ref{fig:ResolvedProf}. This suggests that about $\sim 40\%$ of accreted subhaloes are completely disrupted, and cannot be resolved no matter how much the resolution increases. This disruption fraction is independent of the radial position within the host halo. We will discuss further support to this disrupted fraction when studying the stripping PDF in the Section~\ref{sec:strip}. 


\subsection{The unevolved subhalo mass function}
\begin{figure}
 \myplot{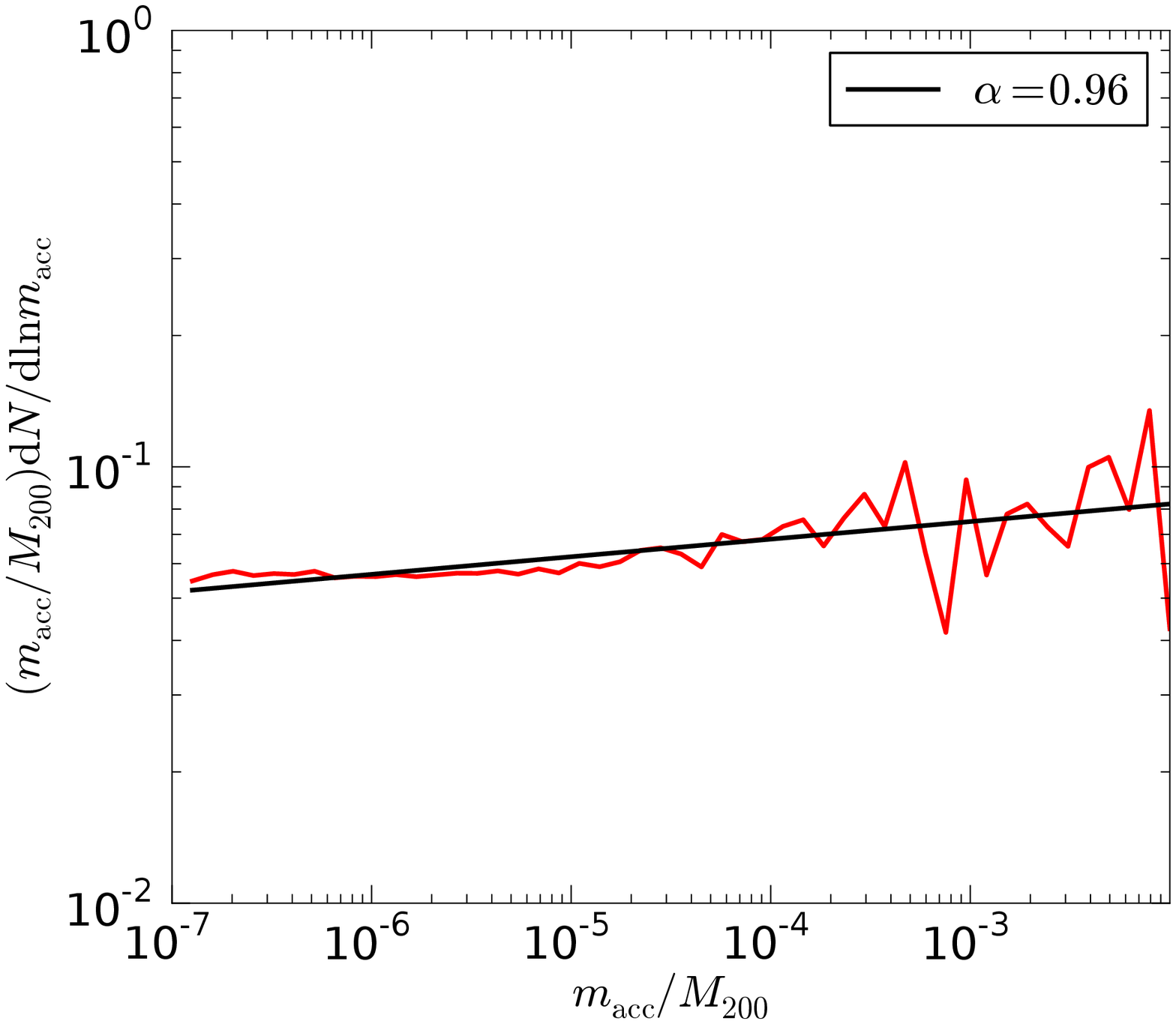}
 \caption{The unevolved subhalo mass function of halo AqA1. The red curve is the data and the solid straight line is a power-law fit in the form of Eq.~\ref{eq:InfallMF}.}\label{fig:InfallMF}
\end{figure}

The abundance of subhaloes accreted by the host over
 all redshifts, at each infall mass $m_{\rm acc}$, is known as the unevolved subhalo mass function~\citep{vdB05}. In principle, this mass function can be obtained semi-analytically from EPS merger trees, or analytically by combining the progenitor mass function predicted by EPS theory with models of a universal halo mass accretion history~\citep{Yang11}, or from EPS alone by considering the environmental dependence of halo formation~\citep{Lee04}. In Fig.~\ref{fig:InfallMF} we show the unevolved subhalo mass function for halo A1, which is well fit by a power law  
\begin{equation}
\frac{\D N}{\D \ln m_{\rm acc}} = A_{\rm acc} \frac{M_{200}}{m_0} \left(\frac{m_{\rm acc}}{m_0}\right)^{-\alpha},\label{eq:InfallMF}
\end{equation}
where $m_0=10^{10}\msunh$ is the mass unit and $A$ and $\alpha$ are parameters. The $M_{200}$ factor is motivated by the scaling of the evolved subhalo mass function~\citep{Gao04a}. For halo AqA1 the best-fitting parameters are $\alpha=0.96$ and $A_{\rm acc}=0.072$. 

\subsection{Tidal stripping of subhaloes}\label{sec:strip}
\subsubsection{\hl{Theoretical motivation}}
Tidal stripping reduces the mass of a subhalo as it orbits inside the host halo. A simple model for the current mass of the subhalo is obtained from the mass inside its tidal radius $r_{\rm t}$. For spherically symmetric density profiles, the instantaneous tidal radius can be obtained by solving the following equation~\citep[see e.g.,][for an equivalent expression]{King}
\begin{equation}
\bar{\rho}(r_{\rm t})=\bar{\rho}(R) \left[2+\kappa_t-\frac{3\rho(R)}{\bar{\rho}(R)}\right],\label{eq:tidal}
\end{equation} where $\bar{\rho}(r_{\rm t})$ is the average density of the subhalo inside $r_{\rm t}$, $R$ is the radial location of the subhalo, $\bar{\rho}(R)$ is the average density of the host halo inside $R$, and $\rho(R)$ is the density of the host at $R$. The parameter $\kappa_t={v_{\rm t}^2}/{V_{\rm c}^2(R)}$ arises due to centrifugal forces in a frame corotating with the subhalo around the host, where $v_{\rm t}$ is the transverse velocity of the subhalo, $V^2_{\rm c}={G M(R)}/{R}$, and $M(R)$ is the host halo mass within radius $R$. Once $r_{\rm t}$ is known, we can calculate the instantaneous bound fraction $\mu=m/m_{\rm acc}$ of the subhalo at $R$, which we name the stripping function $\mu(R)$. In principle, $\mu(R)$ depends on the density profile parameters of both the host and the subhalo. For isothermal density profiles, combined with the virial definition of masses, it is easy to show that $\mu(R)$ only depends on the host halo size through
\begin{equation}
 \mu(R)=\frac{1}{\sqrt{1+\kappa_t}}\frac{R}{R_{200}},
\end{equation}
where $R_{200}$ is the virial radius of the host halo. Equivalent, but more complicated, calculations can be performed for NFW profiles. However for simplicity and motivated 
by the isothermal case, we model the average stripping function of realistic haloes with the approximation
\begin{equation}
 \bar{\mu}(R) = \mu_{\star} \left(\frac{R}{R_{200}}\right)^\beta,\label{eq:stripfunc}
\end{equation} with a slope $\beta \sim 1$, and a normalization $\mu_{\star}$. Note the power-law form of $\bar{\mu}(R)$ is not a requirement for our model. Any other form can be used without affecting our conclusions.

In reality, the stripping of subhaloes is not instantaneous. More importantly, subhaloes are mainly stripped during their pericentric passages and once stripped the mass is not fully regained as the subhaloes move to the outer halo~\citep[e.g.,][]{HBT}. As a result, the current mass of a subhalo also depends on its past trajectory, and can be substantially smaller than the predicted mass inside the instantaneous tidal radius. So the prediction from Eq.~\ref{eq:tidal} should be interpreted as an upper limit to the current mass. Despite this, the average scaling is roughly a scaled version of the upper limit prediction, which, as we will see shortly, can be well approximated by Eq.~\eqref{eq:stripfunc}. Some scatter around the average scaling is also expected, due to the different profiles and trajectories of subhaloes and an evolving host halo profile. The profile of subhaloes can also be modified by tidal heating in addition to truncation. Subhaloes can also fall into the host hierarchically as a subhalo of another subhalo, so that its current mass is also shaped by the tidal field of its host subhalo. All these complexities further add to the scatter around the average stripping function. To be more precise, we will use Eq.~\eqref{eq:stripfunc} to model the median stripping function and model the scatter around it with a log-normal distribution.


For NFW density profiles, it can be shown that a subhalo becomes completely unbound once its size is stripped to below $0.77 r_{\rm s}$~\citep{Hayashi03} \hl{assuming the bound mass does not redistribute after stripping. Motivated by this, } we expect a certain fraction of the accreted subhaloes to be physically disrupted at present, irrespective of the numerical resolution. As a result, the fraction of surviving subhaloes saturates as we move to higher and higher resolution, as was already seen in Fig.~\ref{fig:ResolvedProf}. 

Combining the above arguments, we model the current mass distribution of accreted subhaloes as a mixed distribution of the following form
\begin{align}
 \D P(m|m_{\rm acc}, R)&=(1-f_{\rm s})\delta(m)\D m+ \nonumber\\
 &  f_{\rm s}\mathcal{N}\left(\ln\frac{m}{m_{\rm acc}},\ln\bar{\mu}(R),\sigma\right)\D \ln m. \label{eq:strip_PDF}
\end{align} The first term describes the physically disrupted population, where $f_{\rm s}$ is the survival rate ($1-f_{\rm s}$ is the fraction of physically disrupted subhaloes) and $\delta(x)$ is the Dirac delta function. The second term represents a scaled log-normal distribution, where $\mathcal{N}(x,\bar{x},\sigma)$ is the probability density function of a normal variable of $x$ with mean $\bar{x}$ and standard deviation $\sigma$, and $\bar{\mu}(R)$ is the average stripping function. 
\subsubsection{\hl{Validation of the stripping model}}
\begin{figure}
 \myplot{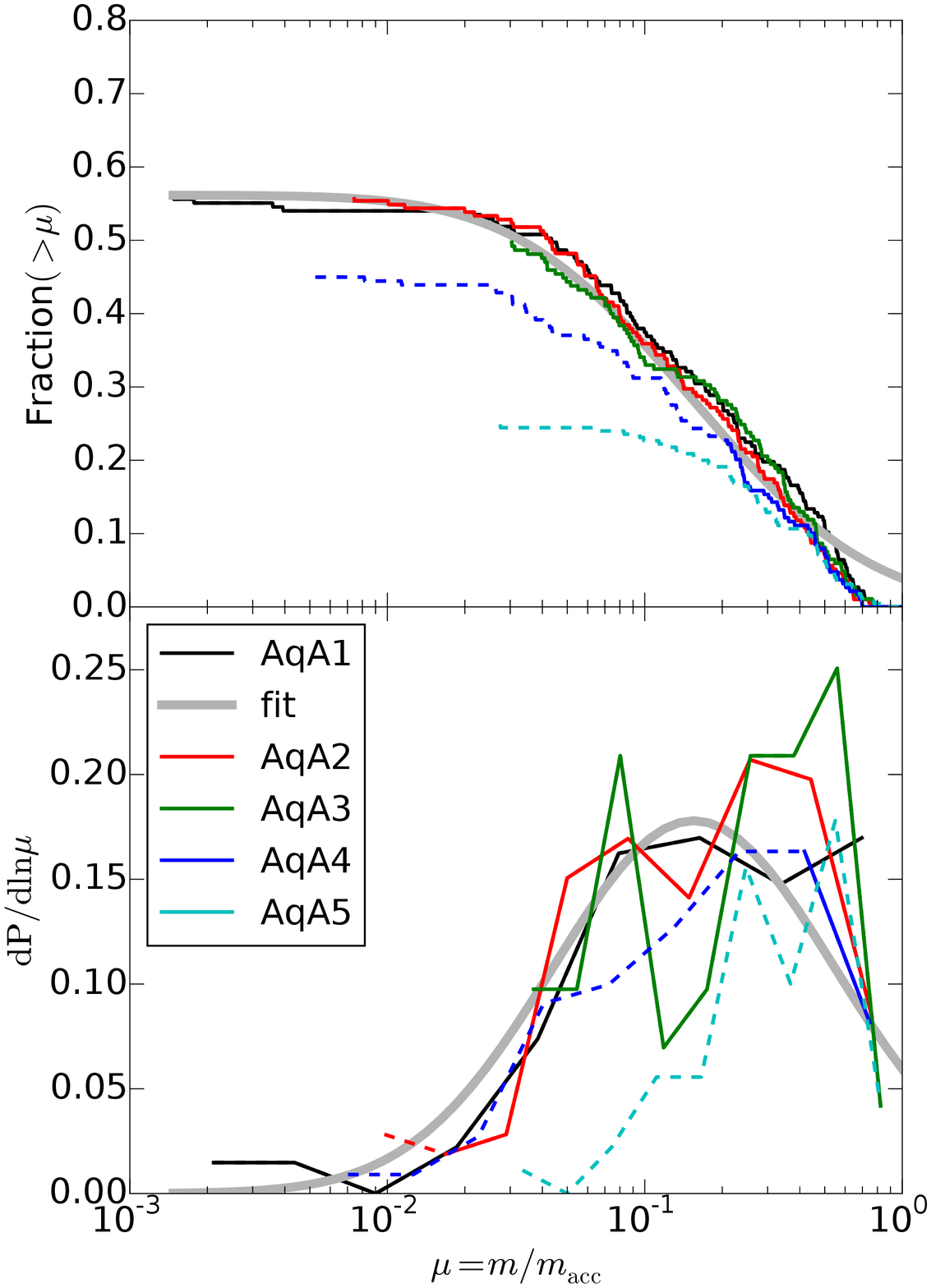}
 \caption{\textit{Top:} the cumulative distribution of $\mu$, for subhaloes in the radial range $0.5-0.8R_{200}$ with infall mass $m_{\rm acc}>10^{-4}M_{200}$. The thick grey line is a fit to AqA1 in the form of Eq.~\eqref{eq:strip_PDF}.  For each line, the solid segment corresponds to where the sample is complete in $\mu$ for subhaloes with more than 100 particles at the present, while the dashed segment refers to the incomplete section. \hl{\textit{Bottom:} the differential version of the top panel, showing the probability density in $\ln\mu$.}}\label{fig:StripCDF}
\end{figure}

In Fig.~\ref{fig:StripCDF}, we show the distribution of surviving  subhaloes at a given radius in the different resolution simulations of halo A. Subhaloes are selected with the same infall mass cut in different simulations, in order to select the same population of objects. \hl{Since subhaloes can only be resolved down to a certain number of particles corresponding to mass $m^{\rm lb}$, the mass fraction sample is only complete down to $m^{\rm lb}/m^{\rm lb}_{\rm acc}$, where $m^{\rm lb}_{\rm acc}$ is the lower limit of the infall mass.} As a result, the completeness limit in $\mu$ differs across simulations of different resolutions, and we separate each line into complete (solid) and incomplete (dashed) portions. Above the sample completeness limit, the distributions from the different resolution simulations agree well. As $\mu$ approaches zero, the cumulative distribution saturates at $f_{\rm s}\approx 0.55$, as expected. The empirical distribution can be well fit by our model distribution (Eq.~\eqref{eq:strip_PDF}), except for the very high $\mu$ end. The deviation from a log-normal distribution at high $\mu$ is expected because $m$ is constrained to be below the instantaneous tidal limit, or at least smaller than $m_{\rm acc}$ by definition. We will revisit this limit when evaluating the evolved subhalo mass function in Section~\ref{sec:fullmodel}. 

\begin{figure}
 \myplot{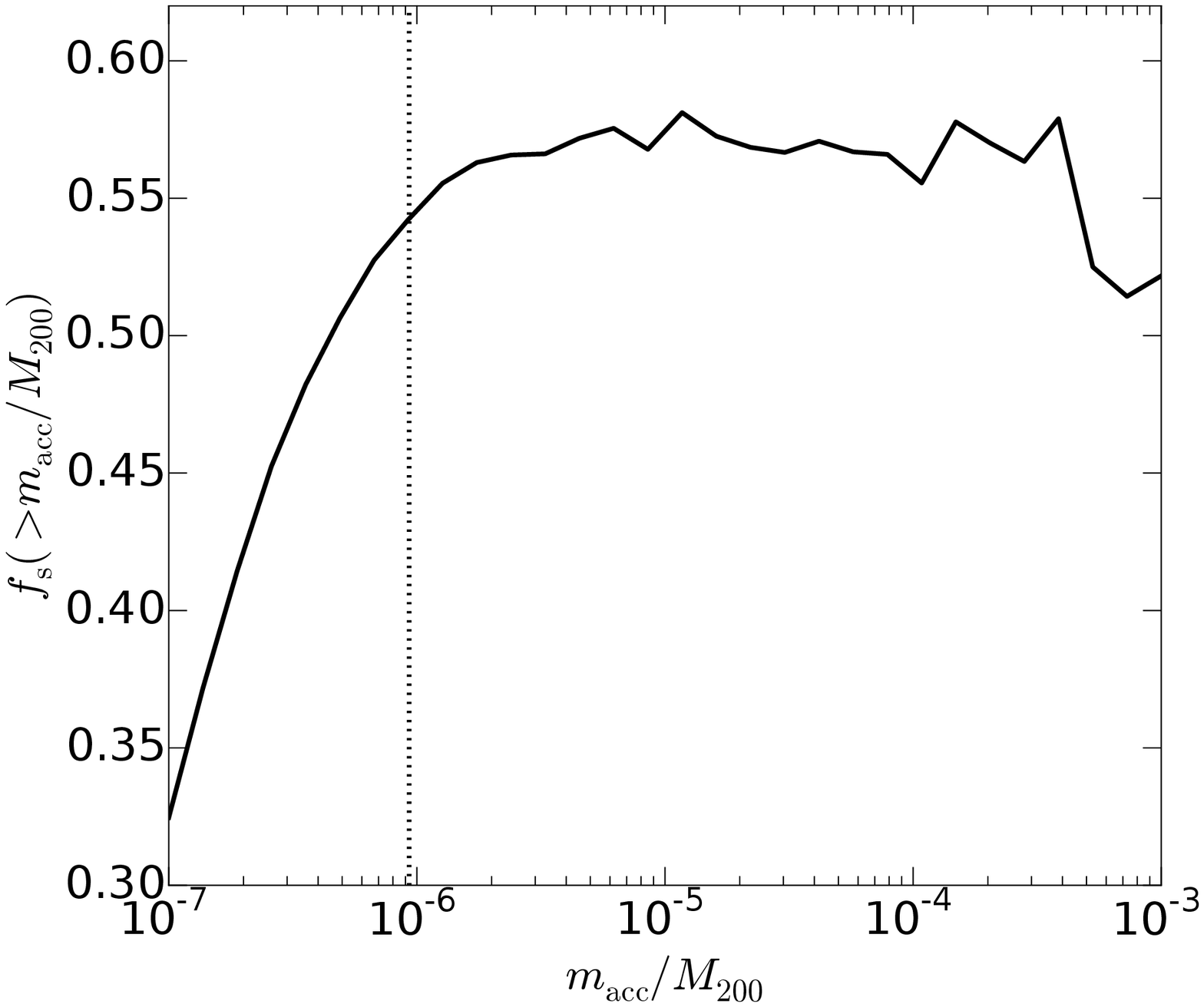}
 \caption{The survival rate $f_{\rm s}$ for subhaloes above a given infall mass, $m_{\rm acc}$, in halo AqA1. The vertical dashed line marks where the infall mass corresponds to 1000 particles, below which the estimate of $f_{\rm s}$ is limited by the incompleteness of the sample.}\label{fig:SurvRateA}
\end{figure}

\hl{In the above model we have assumed the survival rate to be independent of the infall mass. This is examined directly in Fig.~\ref{fig:SurvRateA}. The drop in $f_{\rm s}$ at the low mass end is due to lack of numerical resolution, which limits the ability to resolve the descendants of these small objects. The fluctuation at the very high mass end ($m_{\rm acc}\sim 10^{-3}M_{200}$ and above) is likely to be caused by statistical noise due to the small number of massive subhaloes available. In the dynamical range reliably probed by our simulation, the data are consistent with a constant $f_{\rm s}$.} Note the survival rate is also independent of radius as seen in Fig.~\ref{fig:ResolvedProf}. These two independences further support the approximation that the subhaloes behave like indistinguishable particles in terms of their dynamics.

\begin{figure*}
 \includegraphics[width=0.95\textwidth,height=3.2in]{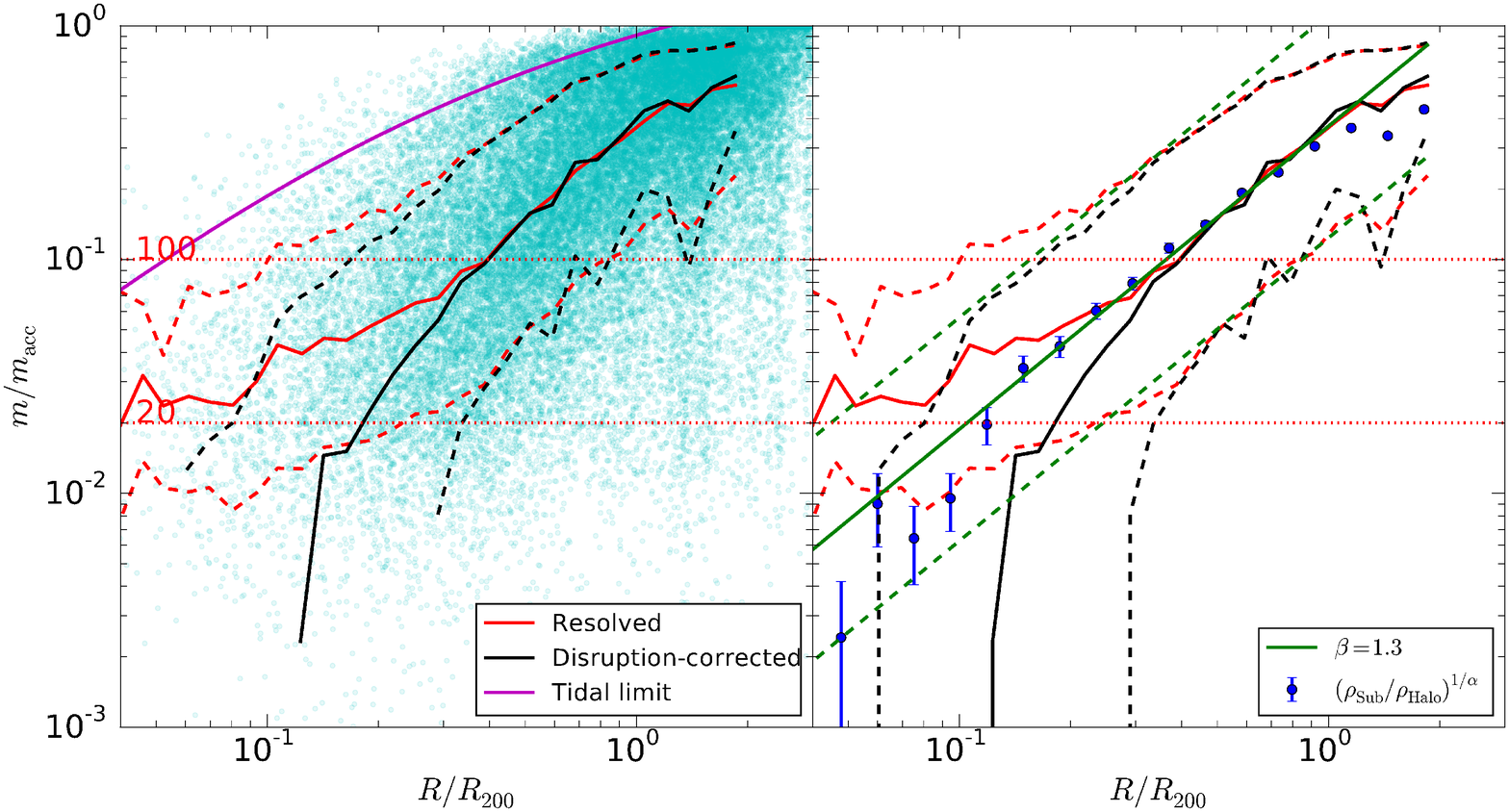}
 \caption{The bound fraction of subhaloes in halo A1. \textit{Left:} The cyan dots in the background show the location and bound fraction of individual subhaloes. The red solid line marks the median bound fraction of resolved subhaloes at each radius while the red dashed lines mark the 16th and 84th percentiles (i.e., $\pm 1\sigma$ confidence intervals). The black lines mark the corresponding percentiles when un-resolved and disrupted subhaloes are also considered. The magenta solid line in the top is the expected bound fraction inside the instantaneous tidal radius for a subhalo inside a MW sized halo according to Eq.~\eqref{eq:tidal}. The red dotted lines mark the limits above which the sample is complete in $\mu$, for subhaloes with more than 20 and 100 particles at the present time respectively. \textit{Right:} \hl{Same as the left panel, but also showing model fits.} The blue circles with errorbars are the $\rho_{\rm Sub}/\rho_{\rm Halo}$ profile in Fig.~\ref{fig:InfallProf} tilted by $1/\alpha$ and scaled to match the normalization of the median bound fraction. The green lines are the fitted median and percentiles of the resolved subhaloes. The black and red lines are identical to those in the left panel.}\label{fig:strip}
\end{figure*}

The radial dependence of the distribution of surviving subhaloes is studied in Fig.~\ref{fig:strip}. \hl{We only include subhaloes with more than 1000 particles ($10^{-6}M_{\rm 200}$) at infall in this figure. As discussed above, the subhalo sample is only complete down to a mass fraction of $m_{\rm lb}/m^{\rm lb}_{\rm acc}$. Two completeness limits are shown, corresponding to a $m^{\rm lb}$ of 20 and 100 particles respectively. }

\hl{A typical instantaneous tidal limit is shown in the left panel. This is calculated from Eq.~\eqref{eq:tidal} for a satellite with an NFW profile at accretion, with a mass of $10^{-4}M_{200}$ and a concentration determined from the average mass-concentration relation of \citep{Ludlow14}. Overall, the tidal limit approximately delineates the upper envelope of the stripped masses, in agreement with expectation. A few data points lie above the tidal limit, because the exact tidal limit depends on the exact density profile of each satellite, which could differ from the one assumed in the calculation. In addition, since tidal stripping is not instantaneous, some subhaloes may not have been subjected to stripping long enough to have lost all the mass outside the tidal radius.}

Two sets of percentiles of the distribution in $\mu$ are extracted at each radius in Fig.~\ref{fig:strip}. First, the median and $\pm 1\sigma$ percentiles are extracted from all the accreted subhaloes that are still resolved at present. However, such percentiles reveal nothing about the disrupted fraction. \hl{Complementing these estimates, we can extract the percentiles using all the accreted subhaloes.} This is possible because even though we do not know the exact $\mu$ value for unresolved or disrupted subhaloes, we do know their abundance at each radius and expect them to lie below the current completeness limit, above which we can still extract percentiles for the full sample. In the presence of disrupted objects, the $p$-th percentile of the surviving population would correspond to a $p'$-th percentile of the full sample where $p'/100=(1-f_{\rm s})+f_{\rm s}p/100$. This is indeed what we see in Fig.~\ref{fig:strip}. \hl{Below the completeness limit, however, the two sets of percentiles diverge. In this regime, the existence of unresolved subhaloes leads to our first method overestimating the percentiles. On the other hand, unresolved subhaloes would cause our second method to underestimate the percentiles. As a result, we expect the true percentiles to lie somewhere in between the two estimates.}

\hl{Before fitting the percentiles, we can first test an important prediction of our model. According to our simplified model (Eq.~\ref{eq:simple_model}), the ratio of the subhalo count profile to host halo density profile, $\rho_{\rm Sub}/\rho_{\rm Halo}$, is directly shaped by the stripping function, $\bar{\mu}(R)$, so that $\rho_{\rm Sub}/\rho_{\rm Halo}\propto \bar{\mu}(R)^\alpha$. We will see later that it is also expected in a more complete model. To test this prediction, we scale $(\rho_{\rm Sub}/\rho_{\rm Halo})^{1/\alpha}$ obtained from Fig.~\ref{fig:prof} to match the amplitude of the median stripping function in Fig.~\ref{fig:strip}. Above the completeness limit, the radial dependence of this ratio matches the two estimates of the stripping function well. Below the completeness limit, it is encouraging to see that the calibrated profile ratio lies safely in between the two estimates, consistent with our expectation of the true stripping function.}

\begin{figure}
 \includegraphics[width=0.49\textwidth,height=3.2in]{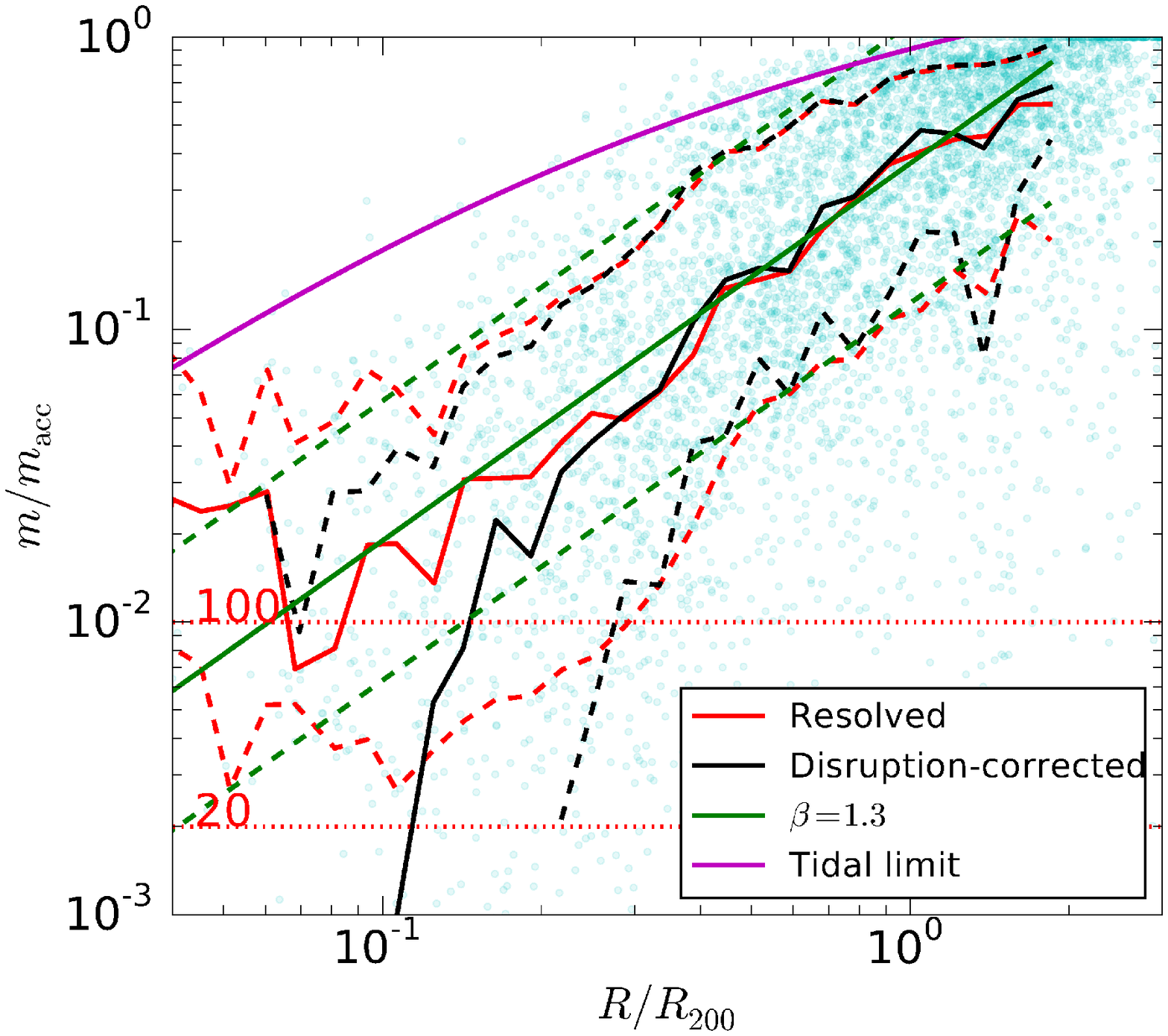}
 \caption{\hl{Same as Fig.~\ref{fig:strip}, but restricted to subhaloes with more than 10000 particles at infall. The green lines (model) are identical to those in Fig.~\ref{fig:strip}.}}\label{fig:strip2}
\end{figure}

\hl{To obtain a parametrized stripping function, we fit a powerlaw to the scaled density profile ratio, $(\rho_{\rm Sub}/\rho_{\rm Halo})^{1/\alpha}$. This fitted line is then shifted by a constant $\sigma_{\ln \mu}=1.1$ vertically which matches well the $\pm 1\sigma$ percentiles. The stripping function is almost independent of the infall mass, as can be seen from Fig.~\ref{fig:strip2} which has a different infall mass selection to that in Fig.~\ref{fig:strip}~\citep[see also][]{Xie15}.}

The above agreements demonstrate that our model adopting a power-law stripping function with a constant log-normal scatter is quite consistent with the current data.

\section{A statistical Model}\label{sec:fullmodel}
The above analysis reveals that our first two assumptions are valid and the third one is also mostly correct except for the scatter around the average stripping function. In the presence of scatter, a more complete treatment starts from the joint distribution of $m$, $m_{\rm acc}$ and $R$:
\begin{equation}
 \D N(m, m_{\rm acc}, R)= \D N(m_{\rm acc}) \tilde{\rho}(R) {\D P(m|m_{\rm acc},R)},\label{eq:joint}
\end{equation} where $\D N(m_{\rm acc})$, $\tilde{\rho}(R)$ and $\D P(m|m_{\rm acc}, R)$ are given by Eqs.~\eqref{eq:InfallMF}, \eqref{eq:rho_def} and \eqref{eq:strip_PDF} respectively. Marginalizing over the infall mass gives the distribution of stripped subhaloes:
\begin{align}
 \frac{\D N(m,R)}{\D \ln m\D^3R} &=A_{\rm acc} \frac{M_{200}}{m_0} \tilde{\rho}(R) \frac{f_{\rm s}}{\sqrt{2\pi}\sigma}\times\nonumber \\
 &\int\displaylimits_{m_{\rm min}}^{m_{\rm max}}  \left[\frac{m_{\rm acc}}{m_0}\right]^{-\alpha} \exp\left[-\frac{1}{2}\left(\frac{\ln \mu-\ln \bar{\mu}(R)}{\sigma}\right)^2\right]\D \ln m_{\rm acc}\nonumber\\
 &= A_{\rm acc} B f_{\rm s} e^{\sigma^2\alpha^2/2} \left[\frac{m}{m_0 \bar{\mu}(R)}\right]^{\alpha} \frac{\rho(R)}{m_0},\label{eq:ModelDistr}
\end{align}
 where $m_{\rm min}$ and $m_{\rm max}$ delimit the mass range of the progenitors that contribute to the final population of  objects of
mass $m$. $B$ is a normalization factor arising from the integral over the log-normal distribution between $m_{\rm min}$ and $m_{\rm max}$:
\begin{align}
 B&=\frac{1}{2}\left[ \erf\left(\frac{\ln (\bar{\mu}/\mu_{\rm min})+\alpha \sigma^2}{\sqrt{2}\sigma}\right)
  -\erf\left(\frac{\ln (\bar{\mu}/\mu_{\rm max})+\alpha \sigma^2}{\sqrt{2}\sigma}\right) \right] \label{eq:B}\\
 &\approx\frac{1}{2}\left[ 1 -\erf\left(\frac{\ln (\bar{\mu}/\mu_{\rm max})+\alpha \sigma^2}{\sqrt{2}\sigma}\right)
 \right], 
\end{align}
 with $\mu_{\rm min}=m/m_{\rm max}\approx m/M_{200}$ and $\mu_{\rm max}=m/m_{\rm min}$.  The second equality holds when $m\ll M_{200}$ so that $\mu_{\rm min}\ll 1$. A reasonable estimate of $\mu_{\rm max}$ is the tidal limit, which is roughly proportional to $\bar{\mu}(R)$ as seen in Fig.~\ref{fig:strip}, making the second term of $B$ a constant. For typical values of $\alpha$ and $\sigma$, adopting $\mu_{\rm max}/\bar{\mu}=5\sim 10$ yields $B \approx 0.7\sim 0.9$. As noted in Section~\ref{sec:strip}, the log-normal distribution does not describe well the high $\mu$ tail of the actual distribution shown in  Fig.~\ref{fig:StripCDF} and so the precise value of $B$ could differ from the above estimate. By matching the predicted subhalo abundance to simulation data, we find a universal value $B\approx0.6$ in both cluster and galaxy haloes, corresponding to an effective $\mu_{\rm max}/\bar{\mu}=4.2$. We fix $B$ to this value hereafter. 

Setting $\sigma=0$ and $f_{\rm s}=1$ recovers the simple model in section~\ref{sec:simplemodel}. When $\sigma\neq 0$, the $\sigma$-dependent term can be understood as arising from Eddington bias in selecting mass $m$ subhaloes from the infall population with $m_{\rm acc}$. This would further modify the radial dependence of the subhalo profile if $\sigma$ depends on $R$. We find that a constant $\sigma\approx1$ is a good approximation to the data. In principle, $f_{\rm s}$ could also be radially dependent. However, because we are mostly interested in the distribution of surviving subhaloes, our model still holds as long as the spatial PDF of surviving subhaloes, $f_{\rm s}\tilde{\rho}(R)$, follows that of the host halo. This is a good approximation as seen in Fig.~\ref{fig:ResolvedProf}. Fig.~\ref{fig:strip} indicates that $f_{\rm s}$ could be larger at smaller radii. This could compensate for the cuspier inner profile of accreted subhaloes in Fig.~\ref{fig:InfallProf}, leading to a profile of surviving subhaloes shadowing that of the dark matter.

\begin{figure}
 \myplot{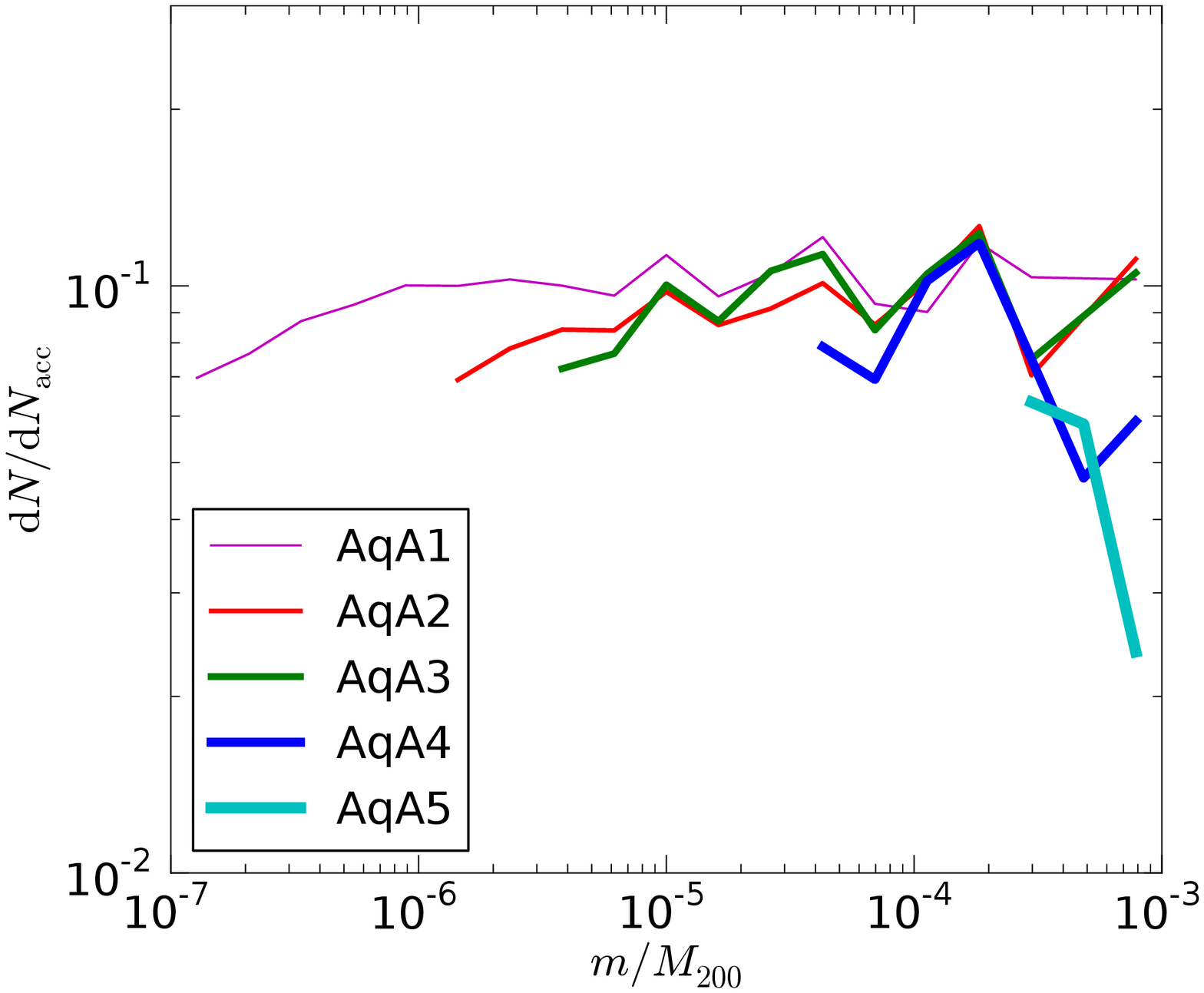}
 \caption{The ratio between un-evolved and final subhalo mass functions, in different resolution simulations of Aquarius halo A.}\label{fig:MFrat}
\end{figure}

This distribution is still separable in $m$ and $R$, leading to a mass-independent spatial distribution of subhaloes as well as a position-independent subhalo mass function. As in the simplified model, the evolved subhalo mass function is predicted to have the same slope as that of the unevolved subhalo mass function. We check for this directly in Fig.~\ref{fig:MFrat}, where we show the ratio between the evolved and unevolved subhalo mass functions. For any given resolution, it appears that the evolved mass function has a shallower slope than the unevolved one at the low mass end. However, this difference is strongly resolution dependent. As the resolution increases, the ratio becomes more and more consistent with a constant over the entire mass range, leading to a conclusion that the underlying fully resolved mass function would have the same slope as that of the un-evolved one. In principle, a small difference can still be introduced by a weak infall mass dependence of $\bar{\mu}(R)$ or a deviation from the log-normal distribution of the stripped fraction $\mu$.  

\begin{figure}
 \myplot{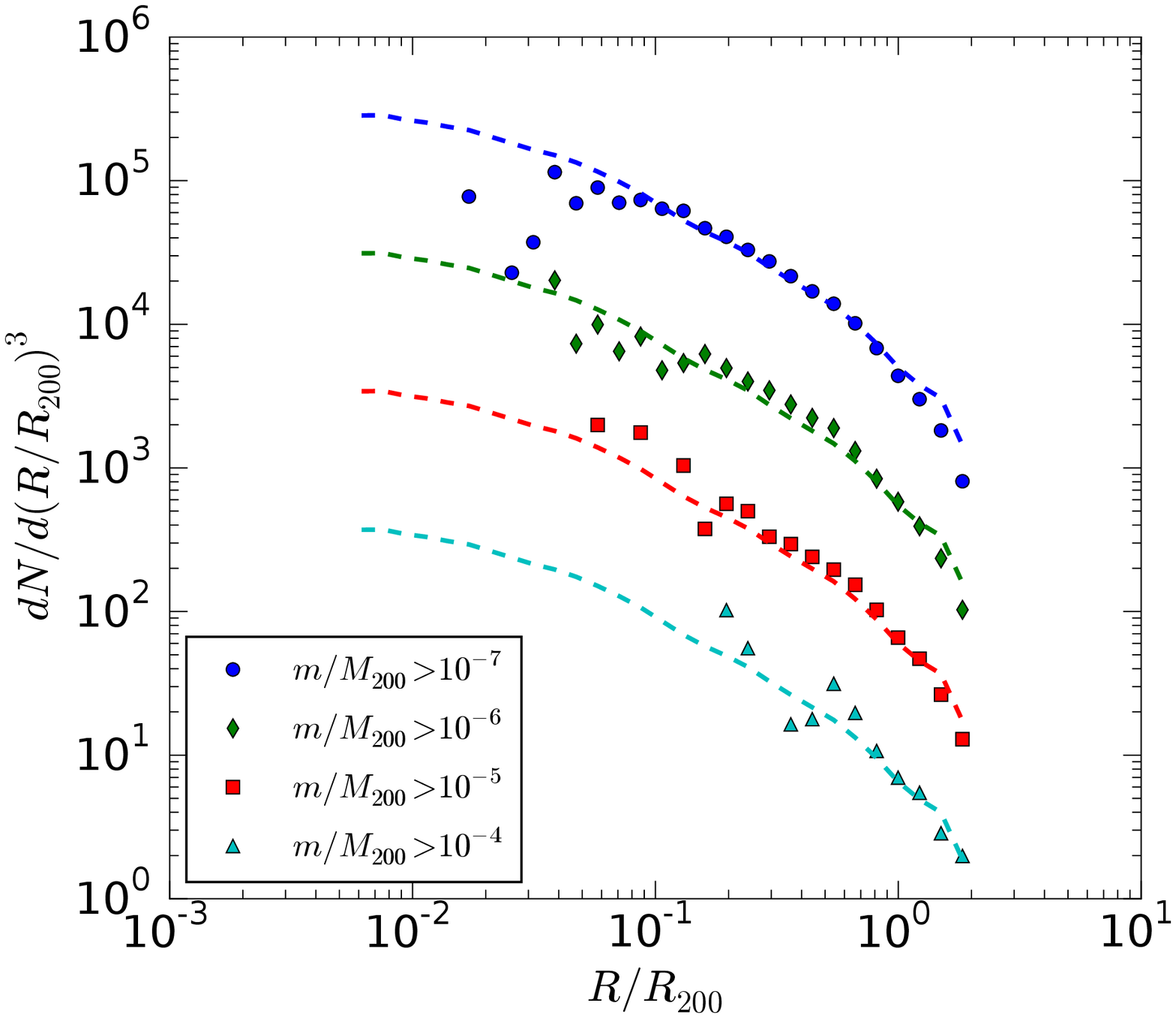}
 \caption{The spatial distribution of subhaloes in halo A1. The points are data from the simulation, selected with increasing lower mass cuts from top to bottom. The dashed lines are model profiles with corresponding cuts.}\label{fig:A1pred}
\end{figure}

In Fig.~\ref{fig:A1pred} we calculate the predicted profile according to Eq.~\eqref{eq:ModelDistr}, with parameters obtained from the previous sections. A good match is found between the model and the data, for different selections in subhalo mass. This means both the subhalo mass function and the subhalo spatial distribution are simultaneously reproduced by the model. 

\subsection{Host mass dependence}\label{sec:calibration}
Focusing on halo A in the Aquarius simulation, we have demonstrated above that a tidal stripping model coupled to a stochastic subhalo profile can describe well the final distribution of subhaloes. In this section we extend the analysis to a sample of haloes from the Aquarius and Phoenix simulations, focusing on the mass dependence of the model. The equations describing the model components apply equally well to these haloes. 
To facilitate general applications of the model, we list in Table~\ref{table:par} the model parameters extracted from these simulations. For the Phoenix simulation, we have only used the first six haloes which are close to each other in mass.  

\begin{table*}
\caption{Parameters of the full model, extracted from the Aquarius (Milky Way) and Phoenix (Galaxy cluster) simulations. For each simulation set, the parameters are estimated from the stacked distribution of the level~2 haloes. 
The ``Joint'' row provides interpolated parameter values intended for application to arbitrary halo masses. $m_0=10^{10}\msunh$ is simply the adopted mass unit. The last row lists the equations in which each parameter is defined.}
\label{table:par}
\begin{center}
\begin{tabular}{ccccccccc}
\hline
\hline  & $M_{200}/(\msunh)$ &$A_{\rm acc}$ & $\alpha$ & $\mu_{\star}$ & $\beta$ & $\sigma$ & $f_{\rm s}$\\
\hline Aquarius & $(1.0 \pm 0.3)\times 10^{12}$ & $0.089$ & $0.95$ & 0.42 & 1.4 & 1.1 & 0.54 \\
\hline Phoenix & $(6.7\pm1.0)\times 10^{14}$ & $0.080$ & $0.95$ & 0.34 & 1.0 & 1.1 & 0.56 \\
\hline Joint &   & $0.1 ({M_{200}}/{m_0})^{-0.02}$ & 0.95 & $0.5 ({M_{200}}/{m_0})^{-0.03}$  & $1.7 ({M_{200}}/{m_0})^{-0.04}$ & 1.1 & 0.55\\ 
\hline
\hline Reference & & Eq.~\eqref{eq:InfallMF} & Eq.~\eqref{eq:InfallMF} & Eq.~\eqref{eq:stripfunc} & Eq.~\eqref{eq:stripfunc} & Eq.~\eqref{eq:strip_PDF} & Eq.~\eqref{eq:strip_PDF}\\
\hline
\end{tabular}			
\end{center}
\end{table*}

The values for most of the parameters are quite similar between the Aquarius and Phoenix haloes. For low precision (e.g., at a level of $\sim 10\%$ in subhalo counts) applications, it is sufficient to simply take the average parameter values as universal and ignore any dependence on halo profile parameters. For applications requiring higher precision, one can interpolate the parameters with respect to halo mass, for $A$, $\mu_{\star}$ and $\beta$, as listed in the ``Joint'' row of Table~\ref{table:par}. We leave more sophisticated calibration of these parameters in different haloes to future work.

\begin{figure}
 \myplot{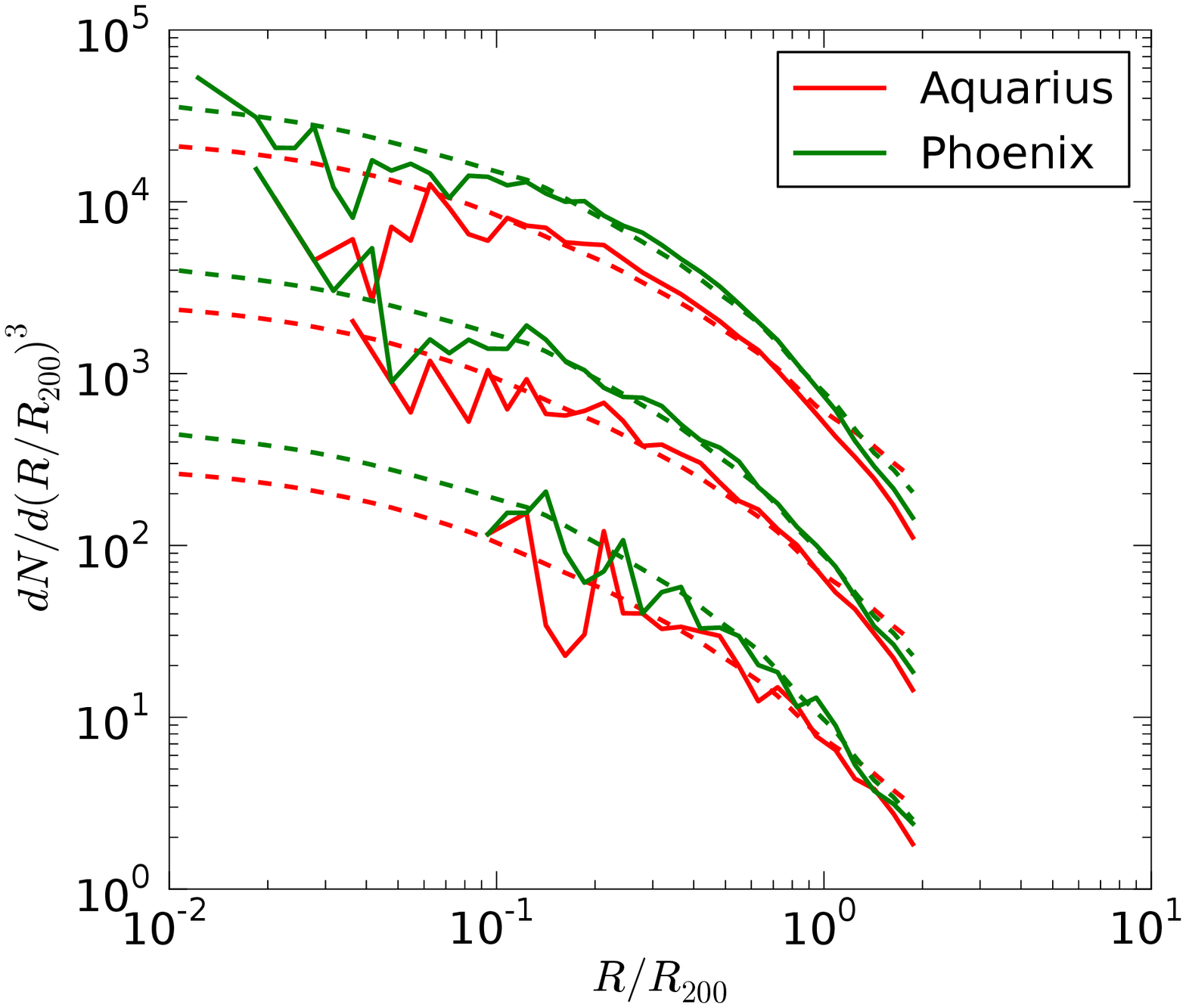}
 \caption{Stacked subhalo profiles. This is analogous to Fig.~\ref{fig:A1pred}, but showing the stacked distribution of subhaloes in the Aquarius and Phoenix simulations. The solid lines are the data and the dashed lines are the model predictions. The three sets of lines correspond to stacking subhaloes with $m/M_{200}$ above $10^{-4}$, $10^{-5}$ and $10^{-6}$ respectively from bottom to top. }\label{fig:ProfPredAll}
\end{figure}

As a consistency check, the stacked subhalo radial distributions in both sets of simulations are shown in Fig.~\ref{fig:ProfPredAll}. As before, the model correctly reproduces the spatial distributions for subhaloes of different masses, in both sets of haloes.

Analytical manipulation of Equation~\eqref{eq:joint} may not always be easy to handle. However, as Eq.~\eqref{eq:joint} is given as a distribution function, it is easy to make Monte-Carlo realizations of subhaloes in the parameter space of $m$, $R$ and $m_{\rm acc}$. Once generated, many other distributions involving these quantities are straightforward to evaluate. For this purpose, we have provided an example \textsc{Python} code, \textsc{SubGen},\footnote{\url{http://kambrian.github.io/SubGen/}} that generates fast subhalo mocks according to Eq.~\eqref{eq:joint}, making use of the \textsc{emcee}~\citep{emcee} Markov-Chain-Monte-Carlo sampler. We will also be taking a Monte-Carlo approach in some of the following applications.

\section{Discussion \& Applications}\label{sec:applications}
The model has important implications for modelling galaxy formation and the statistics of subhaloes in simulations themselves. Once calibrated, the model can be applied to predict many population properties of subhaloes and galaxies. We give some examples below.

\subsection{Do galaxies trace subhaloes?}
The connection between the unevolved and evolved spatial profiles has clear implications for understanding the connection between subhaloes and galaxies. Taking halo-occupation distribution models as an example, an important ingredient in such models is the spatial profile of member galaxies that are used to populate haloes with galaxies. It is common practice to populate haloes with galaxies following the density profile of the host halo. This choice is motivated either due to simplicity by assuming galaxies trace dark matter or by the galaxy distribution in semi-analytical models or SPH simulations \citep[e.g.,][]{Berlind03}. If one assumes galaxies trace subhaloes, however, it appears that the spatial distribution of subhaloes from numerical simulations conflict with this model because the spatial profile of subhaloes appears flatter in the inner halo than that of dark matter. The same discrepancy exists when comparing the observed radial distribution of galaxies with the simulated distribution of subhaloes, a phenomenon termed as the anti-bias of subhaloes. This discrepancy is erased if one recognizes the differences in the two profiles is an effect of sample selection, or an improper comparison between galaxy and subhalo samples. The flatter profile in simulations is a result of selecting in bound mass. In observations, galaxies are more likely to be selected according to stellar mass, which corresponds more closely to selecting subhaloes by infall mass. In this case, planting galaxies following the host halo profile is the correct choice. Such a selection effect has already been noted by \citet{Nagai05,Kravtsov10}. \citet{Conroy06} and \citet{Chen06} showed that populating subhaloes according to their infall properties provides a better match to the observed spatial distribution or clustering of galaxies, although they only studied surviving subhaloes (i.e., no orphan galaxies).

Note the anti-bias is not purely a result of failing to model orphan galaxies. The sample selection bias also operates when only surviving subhaloes are considered. 

\subsection{The ``universal'' subhalo mass function}
\begin{figure}
 \myplot{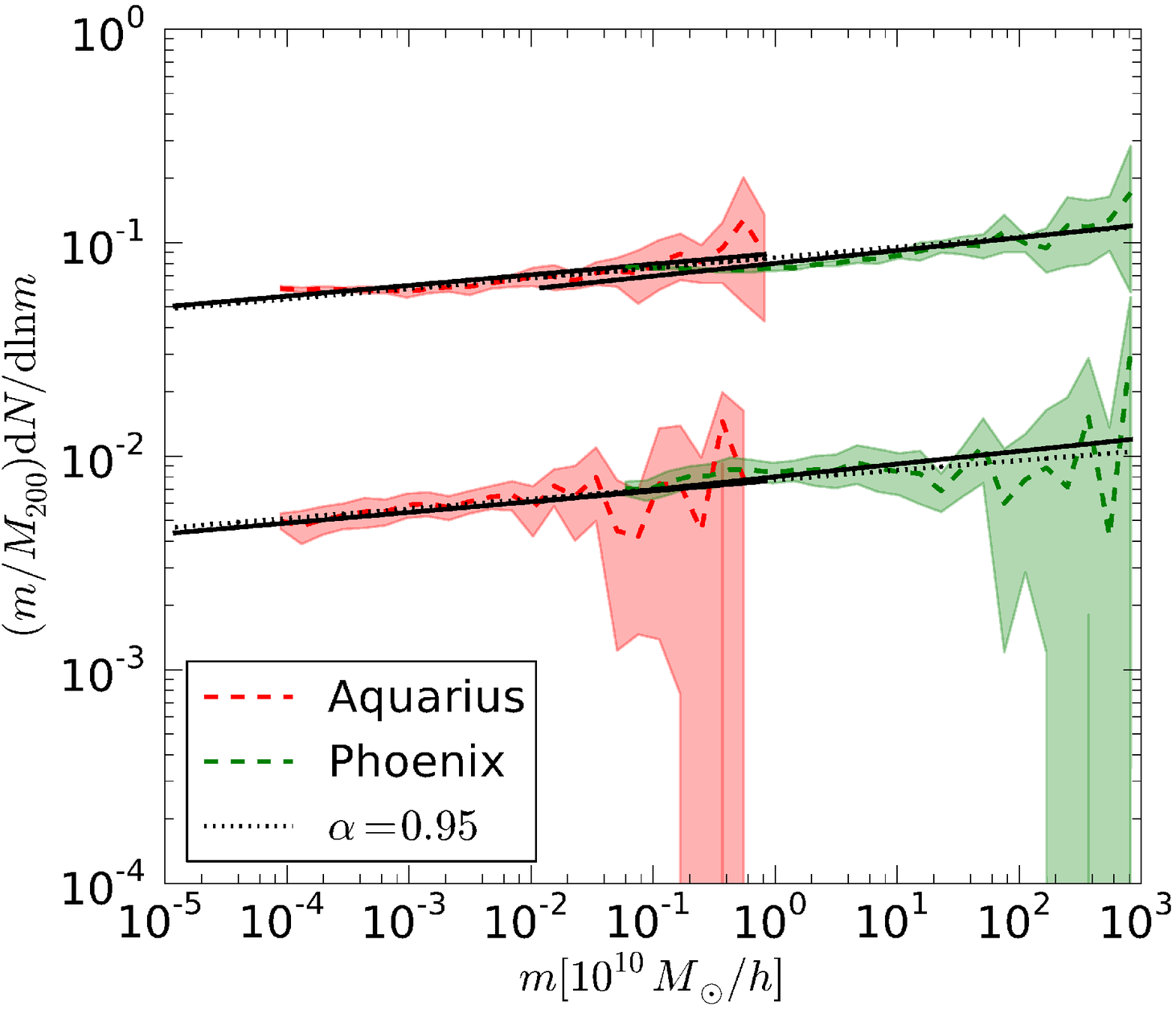}
 \caption{The subhalo mass functions. The shaded regions above and below are the unevolved and evolved subhalo mass functions respectively, with the dashed lines marking the average mass function and the regions spanning the $1$-$\sigma$ scatter. The red and green parts are the results from the Aquarius and Phoenix simulations respectively. The upper two solid black lines are power-law fits (Eq.~\eqref{eq:InfallMF}) to the unevolved mass function with parameters listed in the first two columns of Table~\ref{table:par}. The lower two black solid lines are subsequent model predictions of the evolved mass functions. The dotted lines are joint fits to the two simulations bearing the average parameter values.}\label{fig:MFabs}
\end{figure}

 Subhalo mass functions, both unevolved and evolved, are shown in Fig.~\ref{fig:MFabs}. The evolved subhalo mass function can be obtained by marginalizing over the spatial part of Eq.~\eqref{eq:ModelDistr}, giving
\begin{equation}
 \frac{dN}{d\ln m}=A_{200} \frac{M_{200}}{m_0}\left[\frac{m}{m_0}\right]^{-\alpha},
\end{equation} with
\begin{equation}
 A_{200}=A_{\rm acc} B f_{\rm s} e^{\sigma^2\alpha^2/2}\mu_{\rm star}^\alpha  {\int_0^{R_{\rm 200}} \left(\frac{R}{R_{200}}\right)^{\gamma}\tilde{\rho}(R) \D^3 R}. \label{eq:AApred}
\end{equation} 
Consistent with the good agreement seen in Fig.~\ref{fig:ProfPredAll}, the predicted subhalo mass functions match the data very well. Substituting the parameter values and the numerical halo density profiles into Equation~\eqref{eq:AApred}, we obtain approximately a common normalization parameter $A_{200} \simeq 0.008$ for both sets of simulations, meaning the model automatically reproduces the roughly universal subhalo mass function known in previous studies \citep[e.g.][]{Gao04a}. For comparison, the dotted line in the lower regions show a joint power-law fit to both simulations, which is almost identical to the individual predictions. A combined fit to the unevolved mass function is also shown in the upper region by a dotted line, which unsurprisingly takes the average parameter values listed in Table~\ref{table:par}.

The model predicts an abundance ratio $A_{200}/A_{\rm acc}\simeq 0.1$. It is interesting to note that this is the same as the typical mass fraction locked in subhaloes. The connection between the two would be straightforward if we assume that we can extrapolate the unevolved subhalo mass function down to smooth accretion, $m_{acc}\rightarrow 0$, or if the majority of the halo mass is accreted from subhaloes. In both cases, the host halo mass can be found by adding up contributions from all the progenitors, $M_{200}=\int m_{\rm acc} \D N_{\rm acc}$, so the subhalo mass fraction is simply $\int m\D N/M_{200}=A_{200}/A_{\rm acc}$. We have explicitly checked that in Aquarius A1 the resolved accretion from subhaloes adds up to 90\% of the host halo mass.

\begin{figure}
 \myplot{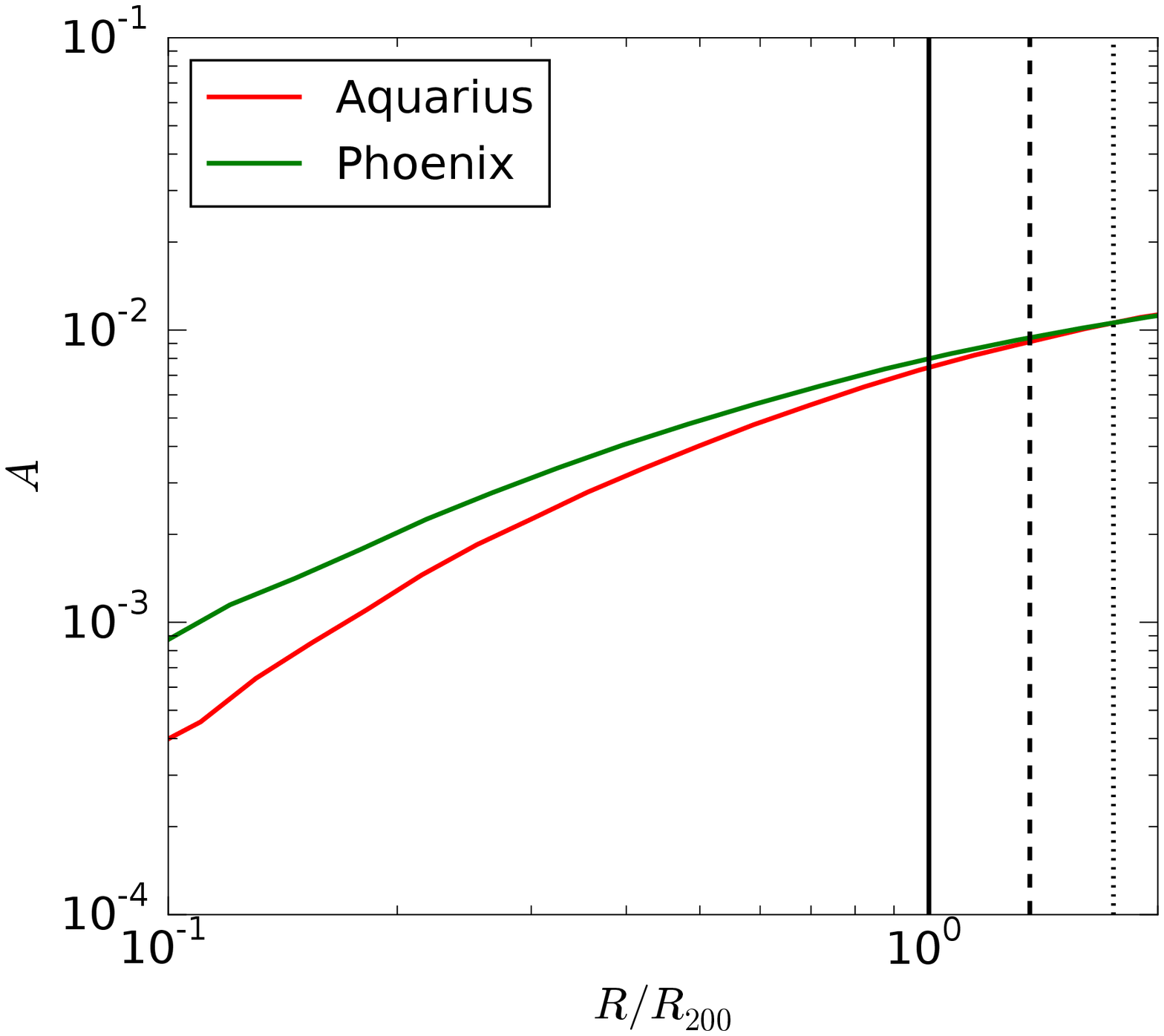}
 \caption{The specific abundance of subhaloes, defined as the normalization of the subhalo mass function in Equation~\eqref{eq:Adef}. The two lines show the average abundances in the Aquarius and Phoenix haloes respectively. The three vertical lines correspond to three common definitions of the virial radius from left to right respectively: $R_{200}$, $R_{\rm vir}$ and $R_{\rm 200b}$ (see text for definitions).}\label{fig:Amplitude}
\end{figure}

The above subhalo mass function is defined to be the abundance of subhaloes inside $R_{200}$. A better understanding of the universality of this mass function can be obtained if we generalize it to be defined inside a different radius, as
\begin{equation}
 \frac{dN(<R)}{d\ln m}=A(R) \frac{M(<R)}{m_0}\left[\frac{m}{m_0}\right]^{-\alpha}.\label{eq:Adef}
\end{equation} The normalization $A(R)$ can be predicted from our model through
\begin{align}
 A(R)=A_{\rm acc} B f_{\rm s} e^{\sigma^2\alpha^2/2} \frac{\int_0^{R} \rho(r) \bar{\mu}(R)^{-\alpha} \D^3 r}{M(<R)}. \label{eq:AApredVir}
\end{align}
In absence of stripping, i.e., for $\bar{\mu}(R)=1$, the subhalo distribution simply follows the host density profile. Then the normalization factor is independent of $R$, and also largely independent of the host halo mass as determined by $A_{\rm acc}f_{\rm s}$. Inside a real halo, however, stripping is important, and the specific abundance of subhaloes would depend on both $R$ and the host halo mass. This is shown in Fig.~\ref{fig:Amplitude}. For $R\gtrsim R_{200}$, tidal stripping is not important. In addition, the majority of the subhalo population inside $R$ is comprised of those close to $R$. This results in the abundance $A(R)$  being largely determined by the infall population, leading to an approximately universal specific abundance $A$ across different halo masses. For $R\ll R_{200}$, however, both a mass and radius dependence is introduced by tidal stripping, because the subhalo profile no longer follows the host profile. The universality of the subhalo mass function thus can be understood as a consequence of the fact that subhaloes trace the density field unbiasedly on large scales ($R>R_{200}$). It is good to see that two other definitions of the virial radius, $R_{\rm vir}$ which is defined to enclose an average density according to the spherical collapse prediction \citep[e.g.,][]{Eke96,BN98}, and $R_{\rm 200b}$ defined to enclose an average density of 200 times the mean matter density, both lie beyond $R_{200}$. Thus the subhalo mass function defined inside $R_{\rm vir}$ and $R_{\rm 200b}$ are also approximately universal. 

Note the specific abundance plotted in Fig.~\ref{fig:Amplitude} is calculated from the simulation data rather than from Equation~\eqref{eq:AApredVir}. This is because Equation~\eqref{eq:AApredVir} is not accurate for $R>>R_{200}$ where the integration over the log-normal distribution in Equation~\eqref{eq:B} breaks down (so that $B$ is no longer a constant) due to the constraint $m/m_{\rm acc}<1$. However, our Monte-Carlo sampler \textsc{SubGen} can handle this boundary condition easily and predicts the right subhalo abundance and spatial distribution even outside $R_{200}$.

\subsection{Dark matter annihilation emission from subhaloes}\label{sec:annihilation}
The dark matter annihilation emission from subhaloes in massive haloes is a very promising target to search for dark matter particles~\citep[e.g.,][]{Gao12,Pinzke11}. With our current model for the distribution of subhaloes, we can proceed to calculate the annihilation emission integrated over the subhalo population. This is achieved by modelling each subhalo with a truncated NFW profile. The NFW parameters are fixed according to the mass and concentration at infall, while the truncation radius can be obtained from the current mass of the subhalo. The relevant equations to obtain the annihilation emission from such a truncated profile are listed in Appendix~\ref{app:tNFW}. If we combine our model with a mass-concentration relation for the subhaloes at infall, we can obtain both the spatial distribution of subhaloes and the truncated density profile of each subhalo. This enables us to evaluate the annihilation emission of each object as well as their distribution inside the host, down to any mass limit of interest. Such a calculation is easy to do with our subhalo sampler, \textsc{SubGen}, and can reproduce the key features of the spatial and mass dependences of the annihilation emission. 

\begin{figure}
 \myplot{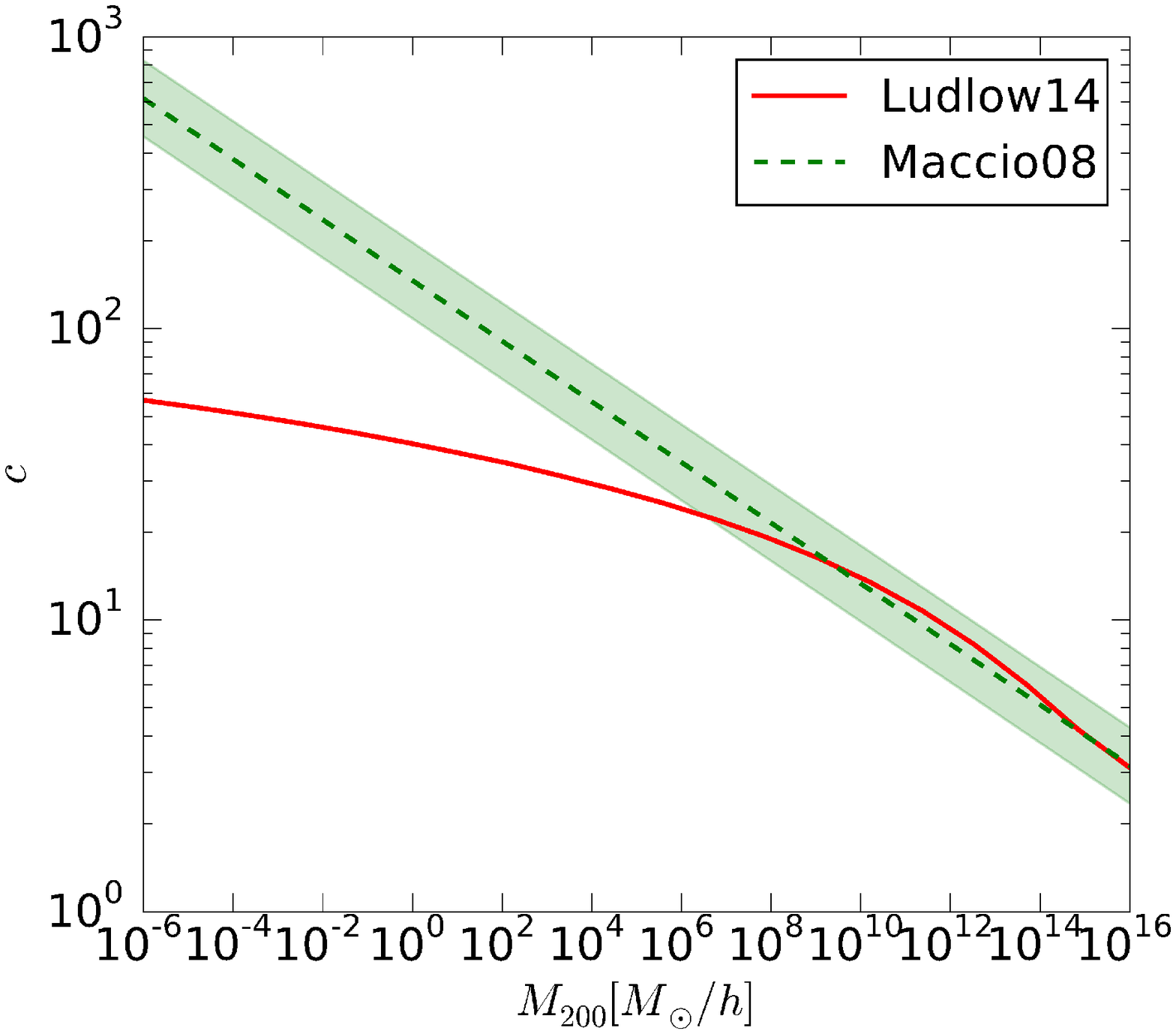}
 \caption{The mass-concentration relations (at redshift 0) used in this work~\citep{Maccio08,Ludlow14}. The green shaded region shows the $\sigma_{\ln c}=0.3$ scatter which we apply to both models.}\label{fig:concentration}
\end{figure}

We assume the concentrations follow a log-normal distribution at fixed mass, with a scatter $\sigma_{\log c}=0.13$. For the median mass-concentration relation, we consider two $z=0$ models from the literature. One is a power-law fit from \citet{Maccio08} for virialized haloes, and the other is a physical model from \citet{Ludlow14} that calculates concentration from the mass accretion history\citep[see also][for some similar models]{Maccio08,Zhao09,Prada12,Correa15}. In both models we adopt the same cosmology as that of our simulations. As shown in Fig.~\ref{fig:concentration}, the two models are consistent with each other for massive haloes resolvable by contemporary cosmological simulations. For small objects ($M_{200}<10^8 \msunh $), however, the \citet{Ludlow14} model predicts a much lower concentration than that extrapolated from a power-law relation.

\begin{figure*}
 \myplottwo{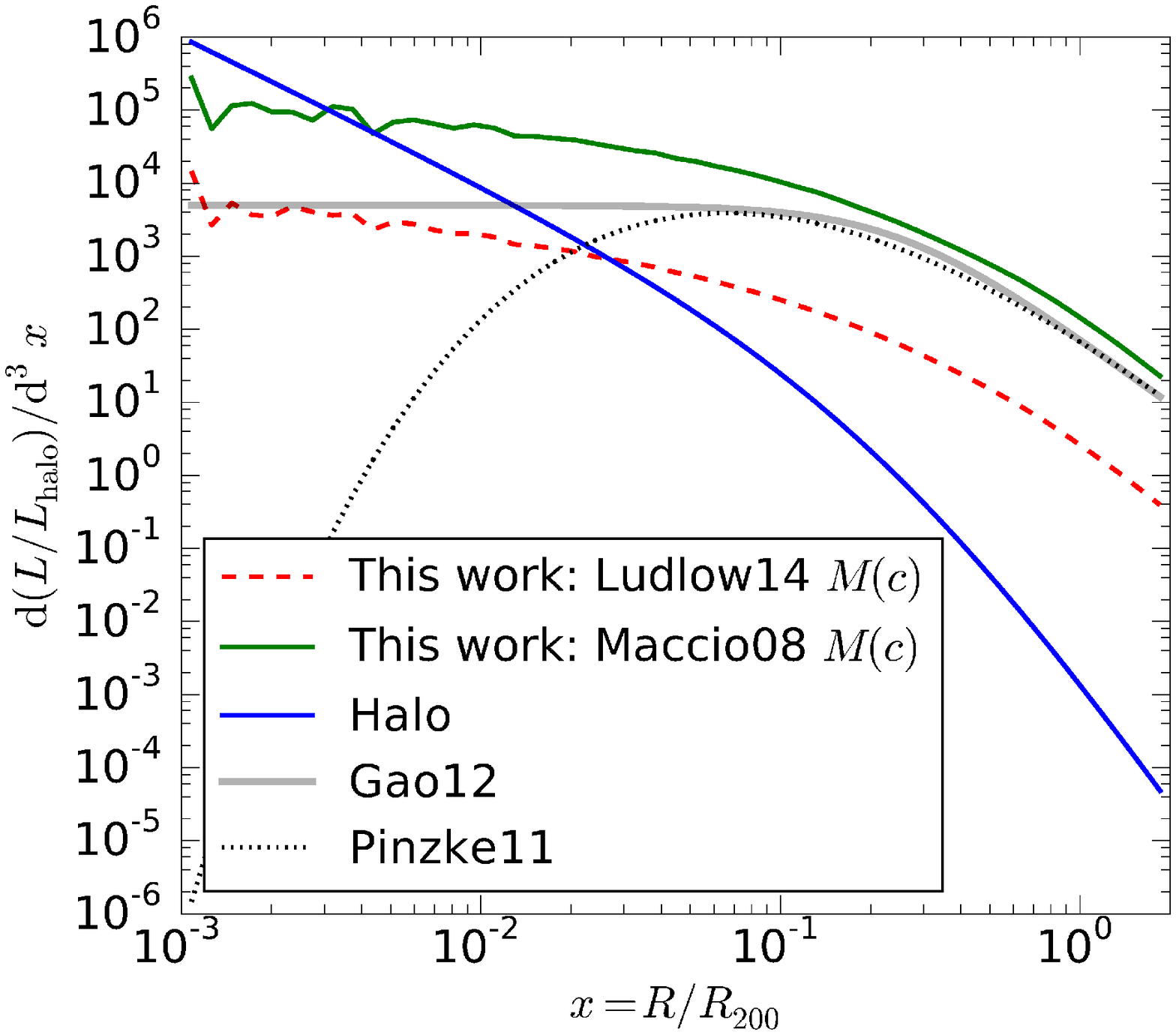}{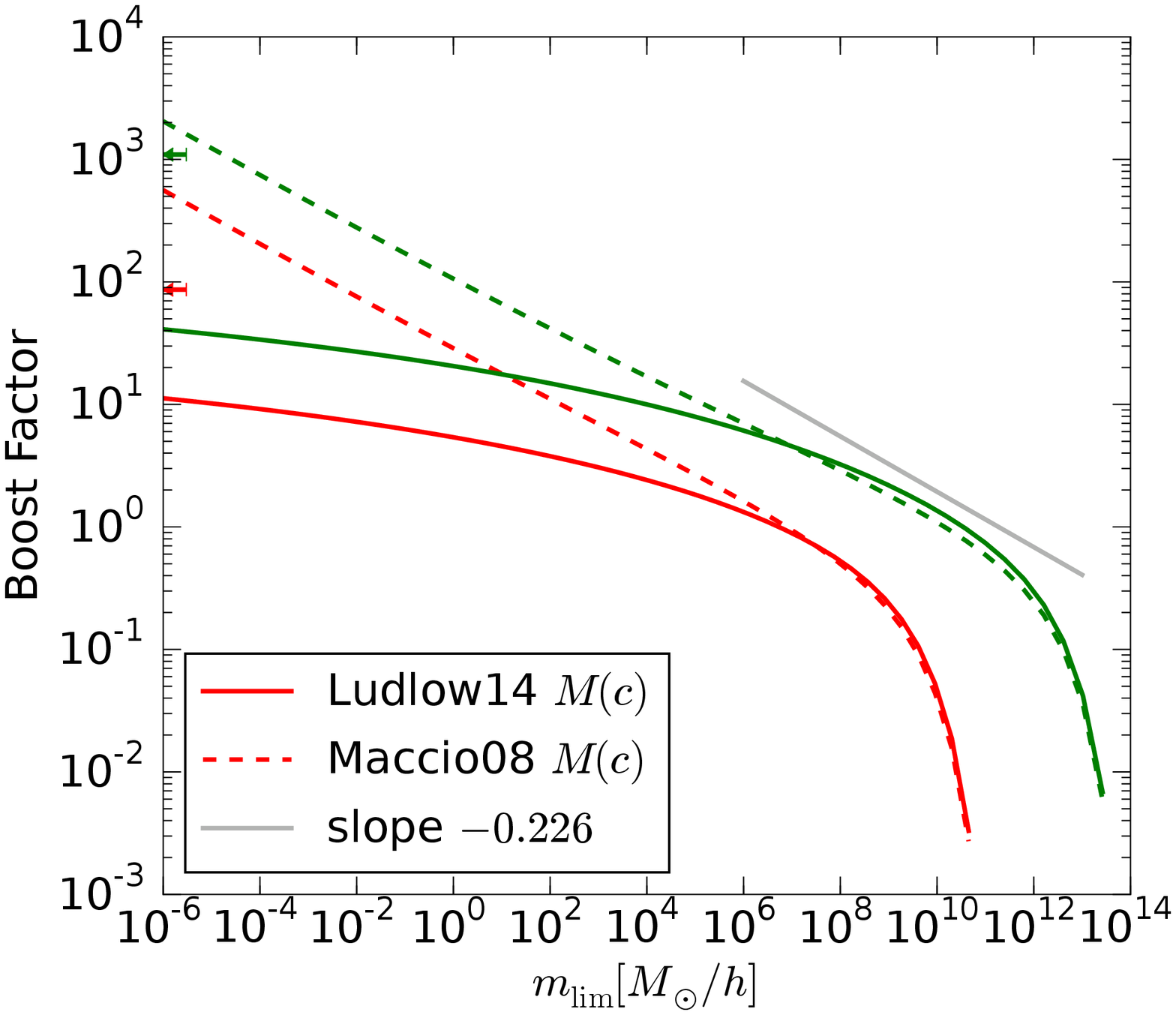}
 \caption{\textit{Left:} The luminosity density from subhaloes in a cluster sized halo ($M_{200}=6.7\times 10^{12}\msunh$). The green solid line and the red dashed line are the predicted annihilation emission from our model adopting the \citet{Ludlow14} and \citet{Maccio08} mass-concentrations respectively. For comparison, the blue solid line show the annihilation emission from the smooth density distribution of the host halo. Two model profiles~\citep{Gao12,Pinzke11} that extrapolate the annihilation emission from resolved subhaloes in simulations are also shown. All the luminosities are normalized by the total luminosity from the smooth density distribution of host halo. \textit{Right:} the boost factor contributed by subhaloes above a given mass limit $m_{\rm lim}$. The red and green lines are for the galaxy and cluster sized halo respectively. For each halo, the dashed line show the result when the a power-law average mass-concentration relation~\citep{Maccio10} is adopted, and the solid line show that when a physical mass-concentration model~\citep{Ludlow14} is adopted. The red and green arrows on the vertical axis mark the extrapolated boost factors for the two haloes according to \citet{Gao12} down to earth mass ($10^{-6}\msun$). The short grey line shows a power-law scaling with $b\propto m_{\rm lim}^{-0.226}$~\citep{Springel08}.}\label{fig:LumDensity}
\end{figure*}

Fig.~\ref{fig:LumDensity} shows the luminosity profile of annihilation emission from subhaloes in a cluster sized halo. For comparison, we also show two models generalized from fitting the Aquarius and Phoenix simulations~\citep{Pinzke11, Gao12}, in normalized coordinates $x=R/R_{200}$ and $\tilde{L}_{\rm sub}=L_{\rm sub}/L_{\rm sub}(R<R_{200})$. Although both profiles are obtained from fitting either Aquarius or Phoenix simulations, the \citet{Gao12} model is fitted to the surface brightness profile (i.e., the projected luminosity profile) of subhalo emission, while the \citet{Pinzke11} model is initially given in the form of the cumulative luminosity profile $L_{\rm sub}(<R)$. The two different ways of fitting and the different parametrizations are likely to lead to disagreement where the profiles are poorly constrained by the simulation. This is evident in Fig.~\ref{fig:LumDensity} for $R\lesssim 0.1R_{200}$. Outside this region, it is encouraging to see the shape of our predicted luminosity profile is in good agreement with both models. Inside this region our model predicts a cuspier profile than the two existing models which are essentially unconstrained by the simulations at such small radii. Adopting the \citet{Ludlow14} profile results in less subhalo emission, while the shape of the luminosity profile from subhaloes is barely affected. 


The amplitude of the subhalo emission is typically specified in terms of a boost factor, defined as the total annihilation luminosity from subhaloes normalized by that from the smooth component of the host halo, both evaluated inside the host virial radius. This is examined directly in the right panel of Fig.~\ref{fig:LumDensity}. Adopting the \citet{Maccio10} mass-concentration relation, the boost factor scales with $m_{\rm lim}$ approximately as a power-law function, with a slope consistent with the value $-0.226$ estimated from the resolved subhaloes in the Aquarius simulation~\citep{Springel08}. Down to an Earth mass, a nominal free-streaming mass scale for cold dark matter haloes, the boost factors rise to a few hundred and a few thousand for galaxy and cluster haloes respectively. These values are slightly higher than those estimated in \citet{Gao12,Springel08} by extrapolating the Aquarius and Phoenix simulations. When the \citet{Ludlow14} mass-concentration relation is adopted, however, the $b(m_{\rm lim})$ function is no longer a power-law, and is significantly reduced at low $m_{\rm lim}$, reflecting the greatly reduced concentration of haloes at the low mass end in this model. Down to an Earth mass, the boost factor is reduced by a factor of 50 in both haloes when using the \citet{Ludlow14} relation compared with that using the \citet{Maccio10} relation. 

The lowered boost factor, in addition to the cuspier emission profile from subhaloes, makes the total luminosity profile inside a halo less extended than that expected from \citet{Gao12,Pinzke11}. This implies that constraints on the dark matter annihilation cross-section in clusters based on previous boost-factor estimates \citep[e.g.,]{Huang12,Han12} could be relaxed. We provide some fitting formulae for the subhalo emission in Appendix~\ref{app:annihilation}.

Our approach differs from some previous estimates \citep[e.g.,][]{Strigari07,Anderhalden13,Sanchez14} in that we start from the infall mass to infer the density profile, rather than from the current mass which has been affected by tidal stripping. The concentration of subhaloes plays a vital role in this estimate, with lower concentrations leading to lower boost factors. We acknowledge several limitations of our current estimate. First, the mass-concentration relation at infall time, instead of that at $z=0$, should be applied to the infall mass. This causes the concentrations to be over-estimated when the $z=0$ relation is used, leading to an over-estimate in the boost factor. For example, lowering the concentration parameters by $0.2$~dex (roughly corresponding to the mass-concentration relation at $z\sim2$) leads to a reduction in the boost factor by a factor of $3$. The correct concentration distribution can be found by looking at the redshift distribution of the progenitors either in simulation or from EPS theory.   We restrain ourselves from calibrating such relations in this work. We notice that \citet{Bartels15} has recently combined analytical models of the unevolved subhalo mass, the accretion-redshift distribution and a redshift dependent mass-concentration with the average mass stripping rate from \citet{Jiang14} for an evaluation of the boost factor. Secondly, the infall mass function is extrapolated to low mass with a power-law form, while in principle it could differ from a power law and could be calculated theoretically with the EPS theory. Thirdly, the stripping function is also assumed to be independent on infall mass down to the lower mass limit of subhaloes. While this is a very good approximation for the subhaloes resolved in our simulations, deviations from a powerlaw form could become important once the infall mass range becomes much larger. A simple estimate utilising  Eq.~\eqref{eq:tidal} suggests a steeper stripping function for low mass subhaloes, which could reduce both the boost factor and the subhalo emission in the inner halo. Despite these limitations, the current model still improves over previous work and can be extended using the current framework. In its current form, our predicted boost factors should be taken as upper limits.

\subsection{The lensing mass profile}
The mass of subhaloes as a function of projected cluster-centric radius at a fixed stellar mass can be measured with weak lensing~\citep[e.g.][]{Li13,Li14,Li15,Sifon15}. Because stellar mass is most directly related to the infall mass of its host subhalo, such measurements essentially constrain the mass of subhaloes selected at fixed infall mass. When stacking subhaloes, disrupted ones (assuming their galaxies persist) contribute no signal, thus they only act to dilute the average signal from surviving subhaloes. In the presence of disrupted subhaloes, the measured signal $\Delta\Sigma_{\rm obs}=f_{\rm s}\Delta \Sigma_{\rm real}$ where $\Delta\Sigma$ is the difference between the cumulative and differential surface density profile of dark matter halo. For subhaloes with (truncated) NFW profiles, the lensing signal $\Delta \Sigma$ is proportional to $m$ (when the other parameters are fixed). Failing to model the disrupted subhaloes would lead to under-estimating the subhalo mass by a factor $f_{\rm s}$. Note this $f_{\rm s}$ is not simply the fraction of orphan galaxies in semi-analytical models, because the latter is not a physical quantity but depends on the resolution of the simulation used by the model. 

Once the disrupted fraction has been corrected for, the measured subhalo mass as a function of projected halo-centric distance can be obtained. Because the surviving mass is not a single value at a fixed infall mass, the measured mass is some average of the underlying mass distribution which in general lies in between the mean and median of the distribution \citep{Mandelbaum05a}. To interpret this ``lensing average'' and to correct it to the true median or mean masses requires knowledge of the underlying mass distribution. For subhalo lensing at a given projected distance, the relevant distribution is
\begin{equation}
 P(m|m_{\rm acc}, R_{\rm p})=\int_{\rm l.o.s} P(\mu|R) \tilde{\rho}(R)\D l.
\end{equation}
Given this distribution, the mean and median subhalo mass can be evaluated analytically or, more conveniently, with the Monte-Carlo sampler \textsc{SubGen}. The generated Monte-Carlo samples can also be used to evaluate the systematic bias in the lensing measurement relative to the real median mass, by simulating the lensing averaging process as was done in \citet{Han14} for real observations. 

\begin{figure}
\myplot{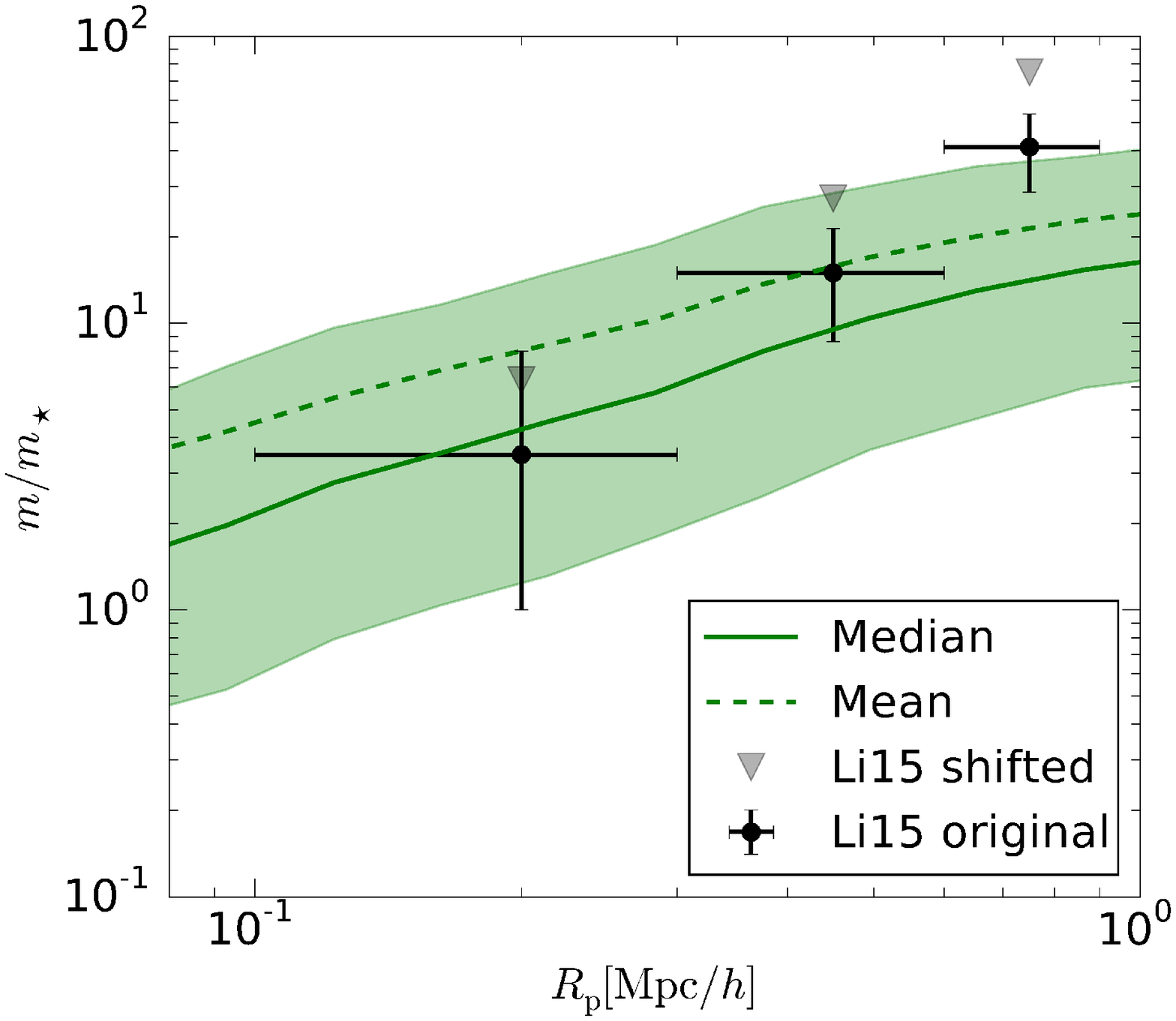}
\caption{The projected profile of subhalo mass to stellar mass ratio in galaxy clusters. The dashed and solid lines represent the mean and median mass of survived subhaloes with stellar mass $M_{\star}>10^{10}\msunh$ in a cluster with $M_{200}=10^{14}\msunh$. The shaded region is bounded by the 15th and 75th percentiles of the sub to stellar mass ratio at each radius. The circles with error-bars are the original measurements from \citet{Li15} while the triangles are original results multiplied by $1/f_{\rm s}$ to account for the disrupted subhaloes.}\label{fig:sublens}
\end{figure}

In Fig.~\ref{fig:sublens}, we compare our predictions with a recent measurement of subhalo mass in galaxy clusters by \citet{Li15}. To populate subhaloes with galaxies (i.e., converting $m_{\rm acc}$ to $m_{\star}$), we adopt the stellar mass-infall mass relation determined in \citet{WangL13} for satellite galaxies with a scatter $\sigma_{\log m_{\star}}=0.19$ at fixed infall mass. We have corrected for the different definitions between our infall mass and that in ~\citet{WangL13}. The subhaloes are further selected with a stellar mass threshold to compare with the observations. Our result is very similar to that from the mock catalogue in \citet{Li15} created from a semi-analytical galaxy formation model, but requires much less effort to obtain. Accounting for the disrupted subhaloes increases the measured mass by a factor of $\sim 2$. After this correction, the measurements start to show a significant tension with the predicted mean and median. However a full investigation which would have to consider many issues is beyond the scope of this paper.
For example, the observational selection function is more complicated than we have assumed and involves selection in host halo mass~\citep{Sifon15} and redshift. Systematic uncertainties in stellar mass estimates may also introduce bias in the mass ratio as well as complicating the selection function. Contamination from neighbouring massive groups is likely to cause an over-estimate at large radii. On the other hand, the value of $f_{\rm s}$ in the real universe may differ from the value used here. Our $f_{\rm s}$ is estimated from dark matter only simulations. The value of $f_{\rm s}$ in the real universe may be different due to baryons, which make subhaloes more resistant to tidal disruption. By applying the $f_{\rm s}$ correction we have also assumed that galaxies are not disrupted together with their host subhaloes, which may not be the case in the real universe. High resolution hydrodynamical simulations are required to address these uncertainties.

\section{Summary \& Conclusions}\label{sec:summary}
We have developed a model that unifies the distribution of subhaloes in mass, $m$, position, $R$, and infall mass, $m_{\rm acc}$. The model fully specifies the joint distribution of these three quantities in an analytical form (i.e. Equation~\ref{eq:joint}):
 \begin{equation}
  \D N(m, m_{\rm acc}, R)= \D N(m_{\rm acc}) \tilde{\rho}(R) {\D P(m|m_{\rm acc},R)},\nonumber
 \end{equation} where $\D N(m_{\rm acc})$ describes the infall mass distribution, $\tilde{\rho}(R)$ is the spatial probability distribution of dark matter particles inside the host halo, and $\D P(m|m_{\rm acc},R)$ describes the final mass distribution of subhaloes of a given infall mass at a given radius. The specific forms of the relevant terms in the joint distribution are given by Equations~\eqref{eq:InfallMF}, \eqref{eq:rho_def} and \eqref{eq:strip_PDF}, with parameter values applicable to different host haloes masses listed in Table~\ref{table:par}. A Monte-Carlo sampler, \textsc{SubGen}, is also provided that easily generates subhalo samples inside any host halo following the above distribution. Once a subhalo sample is generated, any population statistics involving these variables can be easily obtained.

The support for this model can be summarized as follows:
\begin{itemize} 
\item Using high resolution $\Lambda$CDM cosmological simulations of both a galaxy and a cluster sized halo from the Aquarius~\citep{Aquarius} and Phoenix~\citep{Phoenix} projects, we have carefully verified that the shape of the unevolved spatial distribution (i.e., the radial profile at fixed $m_{\rm acc}$) follows the density profile of the host halo, a phenomenon we summarize as \emph{unbiased accretion} of subhaloes. This holds for both surviving subhaloes and unresolved or disrupted subhaloes as traced by their most-bound particles. Dynamical friction leads to a deviation of the unevolved spatial distribution from that of the host halo density profile only in the very inner region and is important only for subhaloes with very large $m_{\rm acc}/M_{200}$. 
\item \hl{The amplitude of the unevolved spatial distribution, as described by the unevolved subhalo mass function, $\D N/\D \ln m_{\rm acc}$, follows a power law in each individual halo.}
\item The joint distribution is then obtained following Bayes theorem, by further specifying the connection between $m$ and $m_{\rm acc}$ with the conditional distribution $P(m|m_{\rm acc},R)$. This connection is shaped by tidal stripping, with subhaloes in the inner halo being more heavily stripped on average. Through a convergence study, we find that about $45\%$ of subhaloes are \emph{physically} disrupted (i.e., stripped to $m=0$ regardless of numerical resolution). Because the spatial distribution is independent of infall mass, the same disruption fraction applies to all infall masses and at all radii. For the surviving subhaloes, we find $P(m|m_{\rm acc}, R)$ can be approximated by a log-normal distribution at each radius, with a median radial dependence well approximated by a power law. 
\end{itemize}
 
Marginalizing (i.e., integrating) the joint distribution over any variable, one obtains the joint distribution of the remaining ones. For example, marginalizing over the infall mass, the model simultaneously reproduces the universal final mass function and the universal spatial distribution of subhaloes of a given final mass. In particular, the model predicts that: 
\begin{itemize}
 \item The final mass function follows a power-law form with the same slope as the infall mass function.
 \item The spatial distribution of subhaloes at fixed $m$, which we call the evolved spatial distribution, is flatter than the density profile of the host halo. The ratio between the two is determined by the amount of tidal stripping at each radius. This explains the so called ``anti-bias'' between the galaxy distribution and the subhalo distribution as purely a selection effect.
 \item The shape of the evolved distribution is also independent of $m$. The scale-free nature (i.e., power-law form) of the infall mass function and the mass-independence of the unevolved spatial profile are the keys to such independence. 
\end{itemize}

The parameters of our model ingredients have been calibrated with simulations and we find only very modest variation with simulated halo mass. This enables the model to be safely interpolated to other halo masses. The calibrated model can be applied to a wide range of problems. We give several such examples, including the  universality of the subhalo mass function, the dark matter annihilation emission from subhaloes, and lensing measurements of subhalo mass. 

We demonstrate that the universality of the subhalo mass function exists because subhaloes trace the density field at large radii where tidal stripping is irrelevant. Further inside this radius, the mass function is lower in more massive haloes. Using the framework to calculate the dark matter annihilation emission of subhaloes, we demonstrate that the adopted mass-concentration relation for subhaloes is crucial in such calculations. Extrapolated down to an Earth mass, the commonly adopted powerlaw mass-concentration model overpredicts the total subhalo emission by a factor of 50 compared with the results when adopting a more physical mass-concentration relation. The model can also be easily adapted to compare with, as well as to calibrate, gravitational lensing measurements of the subhalo mass. The existence of a physically disrupted subhalo population could potentially lead to a correction to the lensing measurement by a factor of $\sim2$, amplifying the tension between a recent subhalo lensing measurement~\citep{Li15} and theoretical predictions.

The model can be extended to higher redshift and further calibrated in other cosmologies. A dependence on host halo concentration may also be introduced as additional model parameters. The aspect of the model that is of the most interest and least known is how subhaloes are stripped. This is described in the model by the average stripping function and its scatter. The unevolved subhalo mass function, on the other hand, can be fully predicted from EPS theory. In addition, EPS calculations are also capable of providing the distribution of accretion redshifts, which can be combined with a redshift-dependent mass-concentration relation to provide accurate density profile parameters for subhaloes. This could for example improve the predictions for the subhalo annihilation emission. 



\section*{Acknowledgments}
We thank Liang Gao and Wojciech Hellwing for helpful discussions, and Aaron Ludlow for providing us a tabulated version of his mass-concentration relation. We thank the anonymous referee for helpful and insightful comments that helped us improve the paper. This work was supported by the Euopean Research Council [GA 267291] COSMIWAY and Science and Technology Facilities Council
Durham Consolidated Grant. YPJ is supported by the 973 program No.~2015CB857003, NFSC~(11533006,11320101002), and a Shanghai key laboratory grant No.~11DZ2260700. This work used the DiRAC Data Centric system at Durham University,
operated by the Institute for Computational Cosmology on behalf of the
STFC DiRAC HPC Facility (www.dirac.ac.uk). This equipment was funded
by BIS National E-infrastructure capital grant ST/K00042X/1, STFC
capital grant ST/H008519/1, and STFC DiRAC Operations grant
ST/K003267/1 and Durham University. DiRAC is part of the National
E-Infrastructure. This work was supported by the Science and
Technology Facilities Council [ST/F001166/1]. 

\bibliographystyle{\mybibstyle}
\setlength{\bibhang}{2.0em}
\setlength\labelwidth{0.0em}
\bibliography{ref}

\begin{thebibliography}{64}
\expandafter\ifx\csname natexlab\endcsname\relax\def\natexlab#1{#1}\fi

\bibitem[{{Anderhalden} \& {Diemand}(2013)}]{Anderhalden13}
{Anderhalden} D., {Diemand} J., 2013, \jcap, 4, 9, \eprint{arXiv:1302.0003}

\bibitem[{{Bardeen} {et~al}\mbox{.}(1986){Bardeen}, {Bond}, {Kaiser}, \&
  {Szalay}}]{BBKS}
{Bardeen} J.~M., {Bond} J.~R., {Kaiser} N., {Szalay} A.~S., 1986, \apj, 304, 15

\bibitem[{{Bartels} \& {Ando}(2015)}]{Bartels15}
{Bartels} R., {Ando} S., 2015, ArXiv e-prints, \eprint{arXiv:1507.08656}

\bibitem[{{Benson} {et~al}\mbox{.}(2002){Benson}, {Lacey}, {Baugh}, {Cole}, \&
  {Frenk}}]{Benson02}
{Benson} A.~J., {Lacey} C.~G., {Baugh} C.~M., {Cole} S., {Frenk} C.~S., 2002,
  \mnras, 333, 156, \eprint{astro-ph/0108217}

\bibitem[{{Berlind} {et~al}\mbox{.}(2003){Berlind}, {Weinberg}, {Benson},
  {Baugh}, {Cole}, {Dav{\'e}}, {Frenk}, {Jenkins}, {Katz}, \&
  {Lacey}}]{Berlind03}
{Berlind} A.~A. {et~al.}, 2003, \apj, 593, 1, \eprint{astro-ph/0212357}

\bibitem[{{Bond} {et~al}\mbox{.}(1991){Bond}, {Cole}, {Efstathiou}, \&
  {Kaiser}}]{Bond91}
{Bond} J.~R., {Cole} S., {Efstathiou} G., {Kaiser} N., 1991, \apj, 379, 440

\bibitem[{{Bryan} \& {Norman}(1998)}]{BN98}
{Bryan} G.~L., {Norman} M.~L., 1998, \apj, 495, 80, \eprint{astro-ph/9710107}

\bibitem[{{Chen} {et~al}\mbox{.}(2006){Chen}, {Kravtsov}, {Prada}, {Sheldon},
  {Klypin}, {Blanton}, {Brinkmann}, \& {Thakar}}]{Chen06}
{Chen} J., {Kravtsov} A.~V., {Prada} F., {Sheldon} E.~S., {Klypin} A.~A.,
  {Blanton} M.~R., {Brinkmann} J., {Thakar} A.~R., 2006, \apj, 647, 86,
  \eprint{astro-ph/0512376}

\bibitem[{{Conroy} {et~al}\mbox{.}(2006){Conroy}, {Wechsler}, \&
  {Kravtsov}}]{Conroy06}
{Conroy} C., {Wechsler} R.~H., {Kravtsov} A.~V., 2006, \apj, 647, 201,
  \eprint{astro-ph/0512234}

\bibitem[{{Correa} {et~al}\mbox{.}(2015){Correa}, {Wyithe}, {Schaye}, \&
  {Duffy}}]{Correa15}
{Correa} C.~A., {Wyithe} J.~S.~B., {Schaye} J., {Duffy} A.~R., 2015, \mnras,
  452, 1217, \eprint{arXiv:1502.00391}

\bibitem[{{Diemand} {et~al}\mbox{.}(2004){Diemand}, {Moore}, \&
  {Stadel}}]{Diemand04}
{Diemand} J., {Moore} B., {Stadel} J., 2004, \mnras, 352, 535,
  \eprint{astro-ph/0402160}

\bibitem[{{Eke} {et~al}\mbox{.}(1996){Eke}, {Cole}, \& {Frenk}}]{Eke96}
{Eke} V.~R., {Cole} S., {Frenk} C.~S., 1996, \mnras, 282, 263,
  \eprint{astro-ph/9601088}

\bibitem[{{Foreman-Mackey} {et~al}\mbox{.}(2013){Foreman-Mackey}, {Hogg},
  {Lang}, \& {Goodman}}]{emcee}
{Foreman-Mackey} D., {Hogg} D.~W., {Lang} D., {Goodman} J., 2013, \pasp, 125,
  306, \eprint{arXiv:1202.3665}

\bibitem[{{Gao} {et~al}\mbox{.}(2004{\natexlab{a}}){Gao}, {De Lucia}, {White},
  \& {Jenkins}}]{Gao04b}
{Gao} L., {De Lucia} G., {White} S.~D.~M., {Jenkins} A., 2004{\natexlab{a}},
  \mnras, 352, L1, \eprint{astro-ph/0405010}

\bibitem[{{Gao} {et~al}\mbox{.}(2012{\natexlab{a}}){Gao}, {Frenk}, {Jenkins},
  {Springel}, \& {White}}]{Gao12}
{Gao} L., {Frenk} C.~S., {Jenkins} A., {Springel} V., {White} S.~D.~M.,
  2012{\natexlab{a}}, \mnras, 419, 1721, \eprint{arXiv:1107.1916}

\bibitem[{{Gao} {et~al}\mbox{.}(2012{\natexlab{b}}){Gao}, {Navarro}, {Frenk},
  {Jenkins}, {Springel}, \& {White}}]{Phoenix}
{Gao} L., {Navarro} J.~F., {Frenk} C.~S., {Jenkins} A., {Springel} V., {White}
  S.~D.~M., 2012{\natexlab{b}}, \mnras, 425, 2169, \eprint{arXiv:1201.1940}

\bibitem[{{Gao} {et~al}\mbox{.}(2004{\natexlab{b}}){Gao}, {White}, {Jenkins},
  {Stoehr}, \& {Springel}}]{Gao04a}
{Gao} L., {White} S.~D.~M., {Jenkins} A., {Stoehr} F., {Springel} V.,
  2004{\natexlab{b}}, \mnras, 355, 819, \eprint{astro-ph/0404589}

\bibitem[{{Giocoli} {et~al}\mbox{.}(2008){Giocoli}, {Tormen}, \& {van den
  Bosch}}]{Giocoli08b}
{Giocoli} C., {Tormen} G., {van den Bosch} F.~C., 2008, \mnras, 386, 2135,
  \eprint{arXiv:0712.1563}

\bibitem[{{Han} {et~al}\mbox{.}(2014){Han}, {Eke}, {Frenk}, {Mandelbaum},
  {Norberg}, {Schneider}, {Peacock}, {Jing}, {Baldry}, {Bland-Hawthorn},
  {Brough}, {Brown}, {Liske}, {Loveday}, \& {Robotham}}]{Han14}
{Han} J. {et~al.}, 2014, ArXiv e-prints, \eprint{arXiv:1404.6828}

\bibitem[{{Han} {et~al}\mbox{.}(2012{\natexlab{a}}){Han}, {Frenk}, {Eke},
  {Gao}, {White}, {Boyarsky}, {Malyshev}, \& {Ruchayskiy}}]{Han12}
{Han} J., {Frenk} C.~S., {Eke} V.~R., {Gao} L., {White} S.~D.~M., {Boyarsky}
  A., {Malyshev} D., {Ruchayskiy} O., 2012{\natexlab{a}}, \mnras, 427, 1651,
  \eprint{arXiv:1207.6749}

\bibitem[{{Han} {et~al}\mbox{.}(2012{\natexlab{b}}){Han}, {Jing}, {Wang}, \&
  {Wang}}]{HBT}
{Han} J., {Jing} Y.~P., {Wang} H., {Wang} W., 2012{\natexlab{b}}, \mnras, 427,
  2437, \eprint{arXiv:1103.2099}

\bibitem[{{Han} {et~al}\mbox{.}(2015){Han}, {Wang}, {Cole}, \& {Frenk}}]{Han15}
{Han} J., {Wang} W., {Cole} S., {Frenk} C.~S., 2015, ArXiv e-prints,
  \eprint{arXiv:1507.00769}

\bibitem[{{Hayashi} {et~al}\mbox{.}(2003){Hayashi}, {Navarro}, {Taylor},
  {Stadel}, \& {Quinn}}]{Hayashi03}
{Hayashi} E., {Navarro} J.~F., {Taylor} J.~E., {Stadel} J., {Quinn} T., 2003,
  \apj, 584, 541, \eprint{astro-ph/0203004}

\bibitem[{{Hellwing} {et~al}\mbox{.}(2015){Hellwing}, {Frenk}, {Cautun},
  {Bose}, {Helly}, {Jenkins}, {Sawala}, \& {Cytowski}}]{CoCo}
{Hellwing} W.~A., {Frenk} C.~S., {Cautun} M., {Bose} S., {Helly} J., {Jenkins}
  A., {Sawala} T., {Cytowski} M., 2015, ArXiv e-prints,
  \eprint{arXiv:1505.06436}

\bibitem[{{Huang} {et~al}\mbox{.}(2012){Huang}, {Vertongen}, \&
  {Weniger}}]{Huang12}
{Huang} X., {Vertongen} G., {Weniger} C., 2012, \jcap, 1, 42,
  \eprint{arXiv:1110.1529}

\bibitem[{{Jiang} \& {van den Bosch}(2014)}]{Jiang14}
{Jiang} F., {van den Bosch} F.~C., 2014, ArXiv e-prints,
  \eprint{arXiv:1403.6827}

\bibitem[{{Jiang} {et~al}\mbox{.}(2015){Jiang}, {Cole}, {Sawala}, \&
  {Frenk}}]{Lilian}
{Jiang} L., {Cole} S., {Sawala} T., {Frenk} C.~S., 2015, \mnras, 448, 1674,
  \eprint{arXiv:1409.1179}

\bibitem[{{Jiang} {et~al}\mbox{.}(2014){Jiang}, {Helly}, {Cole}, \&
  {Frenk}}]{DHalo}
{Jiang} L., {Helly} J.~C., {Cole} S., {Frenk} C.~S., 2014, \mnras, 440, 2115,
  \eprint{arXiv:1311.6649}

\bibitem[{{King}(1962)}]{King}
{King} I., 1962, \aj, 67, 471

\bibitem[{{Kravtsov}(2010)}]{Kravtsov10}
{Kravtsov} A., 2010, Advances in Astronomy, 2010, 8, \eprint{arXiv:0906.3295}

\bibitem[{{Lacey} \& {Cole}(1993)}]{LC93}
{Lacey} C., {Cole} S., 1993, \mnras, 262, 627

\bibitem[{{Lee}(2004)}]{Lee04}
{Lee} J., 2004, \apjl, 604, L73, \eprint{astro-ph/0310853}

\bibitem[{{Li} {et~al}\mbox{.}(2013){Li}, {Mo}, {Fan}, {Yang}, \&
  {Bosch}}]{Li13}
{Li} R., {Mo} H.~J., {Fan} Z., {Yang} X., {Bosch} F.~C.~v.~d., 2013, \mnras,
  430, 3359, \eprint{arXiv:1210.2480}

\bibitem[{{Li} {et~al}\mbox{.}(2015){Li}, {Shan}, {Kneib}, {Mo}, {Rozo},
  {Leauthaud}, {Moustakas}, {Xie}, {Erben}, {Van Waerbeke}, {Makler}, {Rykoff},
  \& {Moraes}}]{Li15}
{Li} R. {et~al.}, 2015, ArXiv e-prints, \eprint{arXiv:1507.01464}

\bibitem[{{Li} {et~al}\mbox{.}(2014){Li}, {Shan}, {Mo}, {Kneib}, {Yang}, {Luo},
  {van den Bosch}, {Erben}, {Moraes}, \& {Makler}}]{Li14}
{Li} R. {et~al.}, 2014, \mnras, 438, 2864, \eprint{arXiv:1311.6523}

\bibitem[{{Libeskind} {et~al}\mbox{.}(2005){Libeskind}, {Frenk}, {Cole},
  {Helly}, {Jenkins}, {Navarro}, \& {Power}}]{Libeskind05}
{Libeskind} N.~I., {Frenk} C.~S., {Cole} S., {Helly} J.~C., {Jenkins} A.,
  {Navarro} J.~F., {Power} C., 2005, \mnras, 363, 146,
  \eprint{astro-ph/0503400}

\bibitem[{{Ludlow} {et~al}\mbox{.}(2014){Ludlow}, {Navarro}, {Angulo},
  {Boylan-Kolchin}, {Springel}, {Frenk}, \& {White}}]{Ludlow14}
{Ludlow} A.~D., {Navarro} J.~F., {Angulo} R.~E., {Boylan-Kolchin} M.,
  {Springel} V., {Frenk} C., {White} S.~D.~M., 2014, \mnras, 441, 378,
  \eprint{arXiv:1312.0945}

\bibitem[{{Macci{\`o}} {et~al}\mbox{.}(2008){Macci{\`o}}, {Dutton}, \& {van den
  Bosch}}]{Maccio08}
{Macci{\`o}} A.~V., {Dutton} A.~A., {van den Bosch} F.~C., 2008, \mnras, 391,
  1940, \eprint{arXiv:0805.1926}

\bibitem[{{Macci{\`o}} {et~al}\mbox{.}(2010){Macci{\`o}}, {Kang}, {Fontanot},
  {Somerville}, {Koposov}, \& {Monaco}}]{Maccio10}
{Macci{\`o}} A.~V., {Kang} X., {Fontanot} F., {Somerville} R.~S., {Koposov} S.,
  {Monaco} P., 2010, \mnras, 402, 1995, \eprint{arXiv:0903.4681}

\bibitem[{{Mandelbaum} {et~al}\mbox{.}(2005){Mandelbaum}, {Tasitsiomi},
  {Seljak}, {Kravtsov}, \& {Wechsler}}]{Mandelbaum05a}
{Mandelbaum} R., {Tasitsiomi} A., {Seljak} U., {Kravtsov} A.~V., {Wechsler}
  R.~H., 2005, \mnras, 362, 1451, \eprint{astro-ph/0410711}

\bibitem[{{Nagai} \& {Kravtsov}(2005)}]{Nagai05}
{Nagai} D., {Kravtsov} A.~V., 2005, \apj, 618, 557, \eprint{astro-ph/0408273}

\bibitem[{{Navarro} {et~al}\mbox{.}(2010){Navarro}, {Ludlow}, {Springel},
  {Wang}, {Vogelsberger}, {White}, {Jenkins}, {Frenk}, \& {Helmi}}]{AqProf}
{Navarro} J.~F. {et~al.}, 2010, \mnras, 402, 21, \eprint{arXiv:0810.1522}

\bibitem[{{Oguri} \& {Lee}(2004)}]{OguriLee04}
{Oguri} M., {Lee} J., 2004, \mnras, 355, 120, \eprint{astro-ph/0401628}

\bibitem[{{Pe{\~n}arrubia} \& {Benson}(2005)}]{PB05}
{Pe{\~n}arrubia} J., {Benson} A.~J., 2005, \mnras, 364, 977,
  \eprint{astro-ph/0412370}

\bibitem[{{Pinzke} {et~al}\mbox{.}(2011){Pinzke}, {Pfrommer}, \&
  {Bergstr{\"o}m}}]{Pinzke11}
{Pinzke} A., {Pfrommer} C., {Bergstr{\"o}m} L., 2011, \prd, 84, 123509,
  \eprint{arXiv:1105.3240}

\bibitem[{{Prada} {et~al}\mbox{.}(2012){Prada}, {Klypin}, {Cuesta},
  {Betancort-Rijo}, \& {Primack}}]{Prada12}
{Prada} F., {Klypin} A.~A., {Cuesta} A.~J., {Betancort-Rijo} J.~E., {Primack}
  J., 2012, \mnras, 423, 3018, \eprint{arXiv:1104.5130}

\bibitem[{{S{\'a}nchez-Conde} \& {Prada}(2014)}]{Sanchez14}
{S{\'a}nchez-Conde} M.~A., {Prada} F., 2014, \mnras, 442, 2271,
  \eprint{arXiv:1312.1729}

\bibitem[{{Sheth} {et~al}\mbox{.}(2001){Sheth}, {Mo}, \& {Tormen}}]{SMT01}
{Sheth} R.~K., {Mo} H.~J., {Tormen} G., 2001, \mnras, 323, 1,
  \eprint{astro-ph/9907024}

\bibitem[{{Sif{\'o}n} {et~al}\mbox{.}(2015){Sif{\'o}n}, {Cacciato}, {Hoekstra},
  {Brouwer}, {van Uitert}, {Viola}, {Baldry}, {Brough}, {Brown}, {Choi},
  {Driver}, {Erben}, {Grado}, {Heymans}, {Hildebrandt}, {Joachimi}, {de Jong},
  {Kuijken}, {McFarland}, {Miller}, {Nakajima}, {Napolitano}, {Norberg},
  {Robotham}, {Schneider}, \& {Verdoes Kleijn}}]{Sifon15}
{Sif{\'o}n} C. {et~al.}, 2015, ArXiv e-prints, \eprint{arXiv:1507.00737}

\bibitem[{{Springel} {et~al}\mbox{.}(2008{\natexlab{a}}){Springel}, {Wang},
  {Vogelsberger}, {Ludlow}, {Jenkins}, {Helmi}, {Navarro}, {Frenk}, \&
  {White}}]{Aquarius}
{Springel} V. {et~al.}, 2008{\natexlab{a}}, \mnras, 391, 1685,
  \eprint{arXiv:0809.0898}

\bibitem[{{Springel} {et~al}\mbox{.}(2008{\natexlab{b}}){Springel}, {White},
  {Frenk}, {Navarro}, {Jenkins}, {Vogelsberger}, {Wang}, {Ludlow}, \&
  {Helmi}}]{Springel08}
{Springel} V. {et~al.}, 2008{\natexlab{b}}, \nat, 456, 73,
  \eprint{arXiv:0809.0894}

\bibitem[{{Springel} {et~al}\mbox{.}(2001){Springel}, {White}, {Tormen}, \&
  {Kauffmann}}]{subfind}
{Springel} V., {White} S.~D.~M., {Tormen} G., {Kauffmann} G., 2001, \mnras,
  328, 726, \eprint{astro-ph/0012055}

\bibitem[{{Strigari} {et~al}\mbox{.}(2007){Strigari}, {Koushiappas}, {Bullock},
  \& {Kaplinghat}}]{Strigari07}
{Strigari} L.~E., {Koushiappas} S.~M., {Bullock} J.~S., {Kaplinghat} M., 2007,
  \prd, 75, 083526, \eprint{astro-ph/0611925}

\bibitem[{{Taylor} \& {Babul}(2001)}]{TB01}
{Taylor} J.~E., {Babul} A., 2001, \apj, 559, 716, \eprint{astro-ph/0012305}

\bibitem[{{Taylor} \& {Babul}(2004)}]{TB04}
{Taylor} J.~E., {Babul} A., 2004, \mnras, 348, 811, \eprint{astro-ph/0301612}

\bibitem[{{Taylor} \& {Babul}(2005{\natexlab{a}})}]{TB05a}
{Taylor} J.~E., {Babul} A., 2005{\natexlab{a}}, \mnras, 364, 515,
  \eprint{astro-ph/0410048}

\bibitem[{{Taylor} \& {Babul}(2005{\natexlab{b}})}]{TB05b}
{Taylor} J.~E., {Babul} A., 2005{\natexlab{b}}, \mnras, 364, 535,
  \eprint{astro-ph/0410049}

\bibitem[{{van den Bosch} \& {Jiang}(2014)}]{vdB14}
{van den Bosch} F.~C., {Jiang} F., 2014, ArXiv e-prints,
  \eprint{arXiv:1403.6835}

\bibitem[{{van den Bosch} {et~al}\mbox{.}(2005){van den Bosch}, {Tormen}, \&
  {Giocoli}}]{vdB05}
{van den Bosch} F.~C., {Tormen} G., {Giocoli} C., 2005, \mnras, 359, 1029,
  \eprint{astro-ph/0409201}

\bibitem[{{Wang} {et~al}\mbox{.}(2013){Wang}, {De Lucia}, \&
  {Weinmann}}]{WangL13}
{Wang} L., {De Lucia} G., {Weinmann} S.~M., 2013, \mnras, 431, 600,
  \eprint{arXiv:1211.4308}

\bibitem[{{Xie} \& {Gao}(2015)}]{Xie15}
{Xie} L., {Gao} L., 2015, ArXiv e-prints, \eprint{arXiv:1501.03171}

\bibitem[{{Yang} {et~al}\mbox{.}(2011){Yang}, {Mo}, {Zhang}, \& {van den
  Bosch}}]{Yang11}
{Yang} X., {Mo} H.~J., {Zhang} Y., {van den Bosch} F.~C., 2011, \apj, 741, 13,
  \eprint{arXiv:1104.1757}

\bibitem[{{Zentner} {et~al}\mbox{.}(2005){Zentner}, {Berlind}, {Bullock},
  {Kravtsov}, \& {Wechsler}}]{Zentner05}
{Zentner} A.~R., {Berlind} A.~A., {Bullock} J.~S., {Kravtsov} A.~V., {Wechsler}
  R.~H., 2005, \apj, 624, 505, \eprint{astro-ph/0411586}

\bibitem[{{Zhao} {et~al}\mbox{.}(2009){Zhao}, {Jing}, {Mo}, \&
  {B{\"o}rner}}]{Zhao09}
{Zhao} D.~H., {Jing} Y.~P., {Mo} H.~J., {B{\"o}rner} G., 2009, \apj, 707, 354,
  \eprint{arXiv:0811.0828}

\end{thebibliography}

\appendix
\section{Annihilation emission from a truncated NFW halo}\label{app:tNFW}
 The mass and concentration parameters of an NFW profile can be easily converted to $(\rho_{\rm s}, r_{\rm s})$, the scale density and scale radius of the halo. If the profile is truncated at $r_{\rm t}$, the truncated mass, $m=m(r_{\rm t})$, can be related to $r_{\rm t}$ through
\begin{align}
 m&=m_{\rm s} \left[\ln (1+x_{\rm t})-\frac{x_{\rm t}}{1+x_{\rm t}}\right]\\
 r_{\rm t}&=-\left[1+\frac{1}{W_0\left(-\mathrm{e}^{-(1+m(r_{\rm t})/m_{\rm s})}\right)}\right]r_{\rm s},\label{eq:NFWrt}
\end{align} 
where $m_{\rm s}=4\pi \rho_{\rm s} r_{\rm s}^3$, $x_{\rm t}=r_{\rm t}/r_{\rm s}$. $W_0(x)$ is the principle branch of the Lambert W function. Subject to a factor that is determined by the particle properties of DM,\footnote{Note our results in Section~\ref{sec:annihilation} and Appendix~\ref{app:annihilation} are always expressed in terms of the normalized luminosity $L/L_{\rm halo}$, which are independent of the particle physics factor.} the annihilation emission of a truncated NFW profile is
\begin{align}
 L(m,m_{\rm acc}, c_{\rm acc})&=\int_{0}^{r_{\rm t}} \rho^2(r) \D^3r\nonumber\\
 &=\left[1-(1+\frac{r_{\rm t}}{r_{\rm s}})^{-3}\right] \frac{4\pi}{3} \rho_{\rm s}^2 r_{\rm s}^3.\label{eq:lum}
\end{align} This luminosity depends very weakly on the truncation radius when $r_{\rm t}>r_{\rm s}$.

\section{Fitting formulas to the subhalo annihilation emission}\label{app:annihilation}
The annihilation emission from subhaloes is usually factorized as
\begin{equation}
 \frac{\D L_{\rm sub}(R)}{\D^3 R}=bL_{\rm host}\frac{\D \tilde{L}_{\rm sub}(x)}{\D^3 x},
\end{equation} where $L_{\rm host}$ is the total emission from the smooth density field of the host halo, $x=R/R_{200}$ is the normalized radius, $\tilde{L}_{\rm sub}(x)=L_{\rm sub}(x)/L_{\rm sub}(x=1)$ is the normalized luminosity with $L_{\rm sub}(x)$ being the total emission from subhaloes inside $x$, and $b$ is the boost factor so that $L_{\rm sub}(x=1)=bL_{\rm host}$.

An analytical function that fits our luminosity profiles reasonably well is 
\begin{equation}
 \frac{\D \tilde{L}_{\rm sub}(x)}{\D^3 x}=0.1(x+0.15)^{-3}.
\end{equation} This profile is not sensitive to the mass of the host halo. It can be further projected to obtain the surface brightness profile for observational applications. The boost factors from subhaloes above an Earth mass can be fitted with
\begin{equation}
 b=4.6\left(\frac{M_{200}}{\msunh}\right)^{0.18}
\end{equation} when adopting the \citet{Maccio10} mass-concentration, and
\begin{equation}
 b=0.08\left(\frac{M_{200}}{\msunh}\right)^{0.18}
\end{equation} when the \citet{Ludlow14} relation is used.

For reference, the luminosity profile and boost factors from \citet{Gao12} and \citet{Pinzke11} are listed below. For the \citet{Gao12} model,
\begin{align}
\frac{\D \tilde{L}_{\rm sub}(x)}{\D^3 x}&=4.53(1+16x^2)^{-3/2}\, ,\label{eq:GaoProf}\\
b&=1.6\times10^{-3}(M_{200}/\msun)^{0.39}\, .
\end{align} Equation~\eqref{eq:GaoProf} is obtained by de-projecting Eq.(2) in ~\citet{Gao12}.\footnote{There is a factor of $2.5$ difference in normalization between Eq.~\eqref{eq:GaoProf} and the original de-projected version. This is because we normalize the profile by the total luminosity inside the 3-D $R<R_{200}$, while ~\citet{Gao12} normalize it by that inside the projected radius $R_{\rm p}< R_{200}$.} For the \citet{Pinzke11} fit,
\begin{align}
\frac{\D \tilde{L}_{\rm sub}(x)}{\D^3 x}&=\frac{a_1[1+ a_2\ln(x)] x^{a_1 x^{a_2}+a_2}}{4\pi x^3},\label{eq:PinzkeProf}\\
b&=0.017\left(\frac{M_{200}}{10^{-6}\msunh}\right)^{0.226},\label{eq:PinzkeBoost}
\end{align} with $a_1=0.95$ and $a_2=-0.27$. Equation~\eqref{eq:PinzkeProf} is obtained from differentiating the original cumulative profile in \citet{Pinzke11} and an earth mass of $10^{-6}\msunh$ has been adopted in Equation~\eqref{eq:PinzkeBoost}.
%
%
%
%
%
%

\end{document}